\documentclass[a4paper,fleqn]{cas-dc}
\usepackage{amsmath,amssymb,amsfonts,amsthm}
\usepackage{textcomp}
\usepackage{xcolor}
\usepackage{graphicx}
\usepackage{cite}
\usepackage{color}
\usepackage{url}
\usepackage{balance}
\usepackage{soul}
\usepackage{enumitem}
\definecolor{darkB}{RGB}{0,51,153}
\setstcolor{darkB}

\newtheorem{assumption}{Assumption}
\usepackage{enumitem}

\newtheorem{Theorem}{Theorem}

\newtheorem{Remark}{Remark}

\usepackage{multicol}
\usepackage{multirow}
\usepackage{caption}
\usepackage{makecell}
\usepackage[subtle]{savetrees}
\usepackage{float}
\usepackage{stfloats}
\usepackage[authoryear,longnamesfirst]{natbib}

\def\tsc#1{\csdef{#1}{\textsc{\lowercase{#1}}\xspace}}
\tsc{WGM}
\tsc{QE}

\begin{document}
\let\WriteBookmarks\relax
\def\floatpagepagefraction{1}
\def\textpagefraction{.001}

\shorttitle{}    

\shortauthors{}  

\title [mode = title]{Stability of DC Microgrids with Nested Nonlinear Control Frameworks: The Fast Communication Scenario}  



%

\author[1]{Cornelia Skaga}

\cormark[1]


\ead{cornelia.skaga@ntnu.no}



\affiliation[1]{organization={Norwegian University of Science and Technology (NTNU)},
            city={Trondheim},
            postcode={7491}, 
            country={Norway}}

\author[2]{Mahdieh S. Sadabadi}

\fnmark[2]

\ead{mahdieh.sadabadi@manchester.ac.uk}



\affiliation[2]{organization={The University of Manchester},
            city={Manchester},
            postcode={M13 9PL}, 
            country={UK}}

\author[3]{Gibert Bergna-Diaz}

\fnmark[3]

\ead{gilbert.bergna@ntnu.no}
\affiliation[3]{organization={Norwegian University of Science and Technology (NTNU)},
            city={Trondheim},
            postcode={7491}, 
            country={Norway}}

\cortext[1]{Corresponding author}


\begin{abstract}
As modern power systems continue to evolve into multi-agent, converter-dominated systems that demand reliable, stable, and optimal control architectures within an expandable framework, this paper investigates \emph{scalable} stability guarantees of a promising \emph{nonlinear} communication-reliant control framework for DC microgrids. Particularly, relying on \emph{nested} control loops--inner decentralized/primary and outer distributed/secondary--the control configurations are designed to simultaneously achieve proportional current sharing and voltage containment within pre-specified limits, at the converged steady state. By enforcing sufficient time-scale separation at the border between the control loops, the system admits a singular perturbation formulation, allowing global exponential stability (G.E.S.) to be established via Lyapunov arguments. Although the theoretical G.E.S. certificate is \emph{structurally scalable}, the stability guarantees depend on a sufficiently large \emph{permanent} leakage--introduced in the primary controller. Thus, this paper emphasizes the importance of appropriate \emph{practical} tuning guidelines for establishing suitable system configurations under which the DC microgrid is \emph{practically} scalable and its equilibrium is G.E.S.. The effectiveness of the proposed method is validated through case studies on a low-voltage DC microgrid under load variations and topological changes, followed by a small-signal stability analysis. 
\end{abstract}

\begin{highlights}
\item Large signal stability guarantees for DC microgrids with modified nonlinear distributed control frameworks. 
\item Employing time-scale separation, singular perturbation theory, and Lyapunov arguments to account for nested primary and secondary control loops, and thus, establishing a global exponential stability certificate.
\item Proposing guidelines to practically guarantee stability, voltage containment, and proportional current sharing in steady state. 
\item Simulation studies are conducted to evaluate the performance of the proposed control framework and to verify that the system converges exponentially to a stable and \emph{near}-optimal steady state, even under topological changes, load variations, or communication disruptions. 
\end{highlights}

\begin{keywords}
DC Microgrids \sep  
Distributed cooperative control \sep 
Nested nonlinear  control loops \sep 
Singular perturbation theory \sep
Lyapunov stability \sep 
Practical control operations \sep 
Current sharing performance
\end{keywords}

\maketitle

\section{Introduction}
In recent years, electric power systems have undergone a significant transition from conventional fossil-fuel-based generation toward the integration of small-scale distributed generation (DG) units such as photovoltaics, wind turbines, and battery energy storage systems, effectively interfaced through power electronic converters (PECs) \cite{Poppi_Budapest, Babak_Set_Point}. Driven by this transformation, modern power grids are evolving into PEC-dominated, multi-agent systems ascribed by their efficient and reliable power management and control strategies \cite{Poppi_Budapest, Babak_AC, CEP_4}. Particularly, given the electrical characteristics of DGs and modern power electronic loads, DC microgrids (MGs)--defined as geographically decentralized, clustered electrical distribution systems--are increasingly recognized as a promising architecture for coordinating multiple converter-interfaced DGs, energy storage systems, and loads within flexible and continuously expandable energy networks \cite{TCNS_2}. 
Moreover, microgrid architectures offer significant operational advantages through their ability to operate in both grid-connected mode--where the microgrid is connected to a larger transmission or distribution system--and autonomous islanded mode, in which the microgrid operates independently under a control architecture that enables coordinated and cooperative operation \cite{TCNS_1, CEP_3}. Such functionality provides a safety mechanism during outages or large disturbances in the utility grid and supports the deployment of DC MGs in industrial applications, including isolated shipboard power systems, marinas, and large ports with flexible loads, aircraft, and electric vehicles \cite{Marine_2, Marine_3, Marine_4, TCNS_2, TCNS_1, CEP_1, CEP_4}.

\subsection{Literature Review and Research Gap}
Recent advances in control theory have shifted control architectures from traditional centralized schemes toward more flexible, scalable, and reliable distributed frameworks that leverage peer-to-peer communication among neighboring units for coordinated decision-making and optimal system operation in accordance with the specified control objectives \cite{Av_Volt_3}. Moreover, the low electrical inertia and weak damping characteristics of PEC-dominated systems increase the vulnerability to transient disturbances and reduce robustness against random perturbations, which further motivates the adoption of efficient and reliable distributed control structures in DC MGs \cite{modeling+stability_rev}. Typically, such distributed architectures follow a hierarchical structure organized into primary, secondary, and tertiary control levels, enabling a certain degree of independence among control layers \cite{Hierarchical}. Within this framework, proper voltage regulation and proportional load (power or current) sharing are commonly defined as the primary control objectives for DC MGs. Voltage regulation is typically enforced through primary PEC controllers--regulating deviations in voltage and/or current (power) outputs relative to a reference value to ensure local voltage stability and reliable network operation \cite{Hierarchical, New_Hier_1, Av_volt_1, Av_Volt_2}. These reference signals are provided by the secondary controller, which commonly employs digital communication links among active agents and consensus-based protocols to cooperatively restore nominal operating conditions; dealing with voltage compensation or power sharing performance, for coordinated operations across the microgrid \cite{Review_Cite_MS_1, Hierarchical, New_Hier_1}. Consequently, in DC MGs--comprising DGs with heterogeneous capacities--the secondary controller enables proportional power sharing, in which the active sources contribute power in proportion to their respective capacities \cite{Av_Volt_5}.

However, without adequate time-scale separation, adverse interactions may arise between the primary and secondary control loops \cite{TCNS_1}. Accordingly, to enable scalable plug-and-play (PnP) functionality and ensure stable, reliable operation of DC MGs--in line with the specified control objectives--while adapting in real time to unknown and variable loads--distributed control structures commonly employ time-scale and bandwidth separation among the primary and secondary control layers \cite{Hierarchical, TCNS_1}.
Recent research has introduced various distributed cooperative control frameworks for DC MGs \cite{cons_DC_1, cons_DC_2, SP_6, cons_DC_5, Av_volt_1, Av_Volt_2, Av_Volt_3}--and references therein--demonstrating the feasibility of achieving simultaneous current sharing and \emph{average voltage regulation}. However, average voltage regulation may lead to significant deviations in individual generator voltages, potentially exceeding operational limits \cite{Babak_Set_Point}. To overcome this limitation, \cite{Babak_Set_Point} introduced a novel nonlinear distributed control framework with \emph{nested} control loops--inner/primary controller and outer/secondary controller. The controller ensures that all energized units contribute power proportionally to their rated capacities while consistently operating within predefined voltage limits. However, the method proposed in \cite{Babak_Set_Point} lacks a rigorous stability guarantee, which serves as the main contribution of this paper.

Furthermore, a closely related control architecture was proposed for AC grids in \cite{Babak_AC}, along with a proof of global asymptotic stability. This proof leveraged singular perturbation theory (SPT) to separate the \emph{nested} control loops by sufficient time-scales, enabling scalable stability conditions under reasonable assumptions for AC systems. In conventional AC MGs, the primary control layer operates at the fastest time-scale with the highest bandwidth, enabling near-instantaneous response to local disturbances, whereas the secondary controller operates at slower time-scales—typically on the order of minutes—with lower bandwidth requirements \cite{Hierarchical, New_Hier_2}. However, in some methods \cite{inverter_1, Inverter_2, Babak_AC} inverters are typically controlled to emulate the droop characteristics of slow synchronous generators \cite{TCNS_1}. Hence, the proof in \cite{Babak_AC} relies on assuming that the inner-loop controller (primary control of the inverters) behaves as decentralized slow virtual synchronous machines, operating at a slower time-scale than the distributed communication-reliant secondary controller.
However, to the best of the authors' knowledge, there are no standardized time-scale requirements for the control levels in DC systems, and given the natural \emph{structural} differences between AC and DC grids, assuming similar time-scale separations may arguably be counter-intuitive. Hence, it is not entirely obvious that the stability conditions in \cite{Babak_AC} still hold \emph{mutatis mutandis} in the DC grid case in \cite{Babak_Set_Point}.  

\subsection{Contribution}
The primary contribution of this paper is the derivation of a \emph{scalable} global exponential stability (G.E.S.) certificate for closed-loop DC microgrids with \emph{nonlinear nested} control architectures. To this end, the paper first introduces the notion of \emph{scalability} in Remark~\ref{rem_stuct_pract}, distinguishing between \emph{structurally} and \emph{practically scalable} stability guarantees.
\begin{Remark}\emph{(Scalable Stability Definitions)} 
\label{rem_stuct_pract} 
\begin{itemize}
    \item \textbf{Structurally scalable} refers to the property for which the G.E.S. certificate remains applicable for any scale of the system, thereby facilitating expandable system architectures without altering the underlying form or structure of the stability proof.
    \item \textbf{Practically scalable} refers to the property for which the G.E.S. certificate remains valid for expandable systems, given that the proposed system configurations guarantee stability for the most critical operating point. Consequently, stable PnP properties are only guaranteed within a certain range of operating conditions and parameter settings.
\end{itemize}
\end{Remark}
Subsequently, motivated by the stability analysis presented in \cite{Babak_AC} for closed-loop AC microgrids, this paper aims to establish \emph{structurally scalable} G.E.S. guarantees for the closed-loop DC microgrid. In addition, \emph{practical} guidelines are proposed to establish system configurations (controller tunings and electrical specifications) that satisfy the stability conditions at the most critical operating point, thereby achieving \emph{practical scalability}.

Lyapunov’s direct method -- commonly used in large-signal stability analysis of nonlinear systems-- can indeed be used to assess the stability of the case-specific system with a defined (and preferably low) number of generators admitting the structure given in \cite{Babak_Set_Point}. However, the introduction of \emph{nonlinear} dynamics together with \emph{nested} control loops significantly complicates the derivation of global exponential G.E.S. certificates within an expandable framework. Accordingly, the method adopted in this paper follows the methodology employed in \cite{Babak_AC}; that is, enforcing time-scale separation between the fast communication-based outer-loop (secondary) control dynamics and the slower inner-loop (primary) control and electrical dynamics. Under this separation, G.E.S. guarantees can be established for the two time-separated subsystems, thereby ensuring that the overall closed-loop DC microgrid is G.E.S.

Moreover, building upon the preliminary presented stability analysis in \cite{Poppi_Budapest}, this paper advances the technical results in the following ways: (i) we update the control dynamics to include a nonlinear and \emph{non-permanent} leakage function of the primary inner-loop controller--for improved dynamical performance--and a \emph{permanent} leakage to facilitate the derivation of a \emph{structurally scalable} G.E.S. certificate, (ii) the stability proof has been substantially updated to accommodate these control modifications, and to guarantee G.E.S. in both saturated and unsaturated operations, (iii) we propose some \emph{practical} guidelines to enforce \emph{practically scalable} stability guarantees. Furthermore, the case studies are extended to empirically validate the theoretical stability result for systems operating under the \emph{practical} guidelines, and to evaluate whether the proposed controller consistently ensures voltage containment and convergence to a steady state ensuring \emph{practical} proportional current sharing. Given that SPT results in, to some extent, conservative stability guarantees due to the critical time-scale separation requirements, the stability of the system is further examined under relaxed time-scale separation between the nested control loops. Specifically, (i) we impose communication disruptions that both satisfy and violate the required time-scales of the outer-loop controller, and (ii) a small-signal stability analysis is performed via a parametric sweep of the outer-loop time constants. Together, these analyses assess the potential to reduce the conservativeness of the large-signal stability result, thereby extending its area of application.

The rest of the paper is structured as follows. Section \ref{Model} presents the dynamical model of the electrical system and distributed control framework for a DC MG. In Section \ref{Stability}, we present the main result of this paper: a comprehensive stability proof for an $n$-generator closed-loop DC microgrid with a nonlinear distributed controller that under certain \emph{practical} guidelines can stabilize to a desired steady state. Furthermore, in Section \ref{sec_Case_Studies}, the controller is evaluated via time-domain simulations on a 4-terminal low-voltage DC microgrid to demonstrate the effectiveness of the proposed method, including testing the performance under communication disruptions, followed by a small-signal stability analysis to complement the large-signal stability results. Finally, Section \ref{conc} concludes this paper.  

\section{System Modeling and Control Framework} \label{Model}
Throughout the paper, we use the following notations: $\mathbb{R}^{n\times m}$ and $\mathbb{R}^n$ denote a set of $n\times m$ real matrices and $n \times 1$ real vectors, respectively. $\mathrm{col}(\cdot\cdot\cdot) \in \mathbb{R}^n$ denotes a column vector and $\mathrm{bcol}\{\cdot\cdot\cdot\}\in \mathbb{R}^n$ denotes a column vector of vectors. $\mathrm{diag}(\cdot\cdot\cdot)\in \mathbb{R}^{n\times n}$ denotes a diagonal matrix with scalar entries, and $\mathrm{bdiag}\{\cdot\cdot\cdot\}\in \mathbb{R}^{n\times n}$ denotes a block diagonal matrix of vectors. $0$ denotes a null matrix of appropriate dimensions, and $\mathbb{I}=\mathrm{diag}(1)$ denotes the identity matrix of appropriate dimensions. $A|_\mathrm{sym}$ denotes the symmetrical part of an arbitrary square matrix $A \in \mathbb{R}^{n \times n}$, defined as $A|_\mathrm{sym}=\frac{1}{2}(A+A^\top)$. Given a symmetric matrix $X$, $X \succ 0$ ($X \geq 0$) indicates that $X$ is a positive-definite (positive semi-definite) matrix. Furthermore, $x \gg y$ denotes that x is much larger than y, and $\mathbb{R}_{\geq 0}$ denotes all non-negative real values. Given a scalar or a vector $x$, the value at the equilibrium point is indicated as $\Bar{x}$, and $\tilde x$ denotes a shifted variable where $\tilde x \triangleq x - \bar x$. Note that $\mathcal{G}=\{1, 2, \cdot \cdot \cdot, n_i\}$ is the set of distributed generators, $\mathcal{E}=\{1, 2, \cdot \cdot \cdot, n_j\}$ is the set of power lines, $\mathcal{N}=\{1, 2, \cdot \cdot \cdot, n_k\}$ is the set of power consuming loads, and $\mathcal{N}^c=\{1, 2, \cdot \cdot \cdot, n_i\}$ is the set of neighboring communicating DGs.

\subsection{Dynamical Model of Microgrids}
\begin{figure}
    \centering
    \includegraphics[width=\columnwidth]{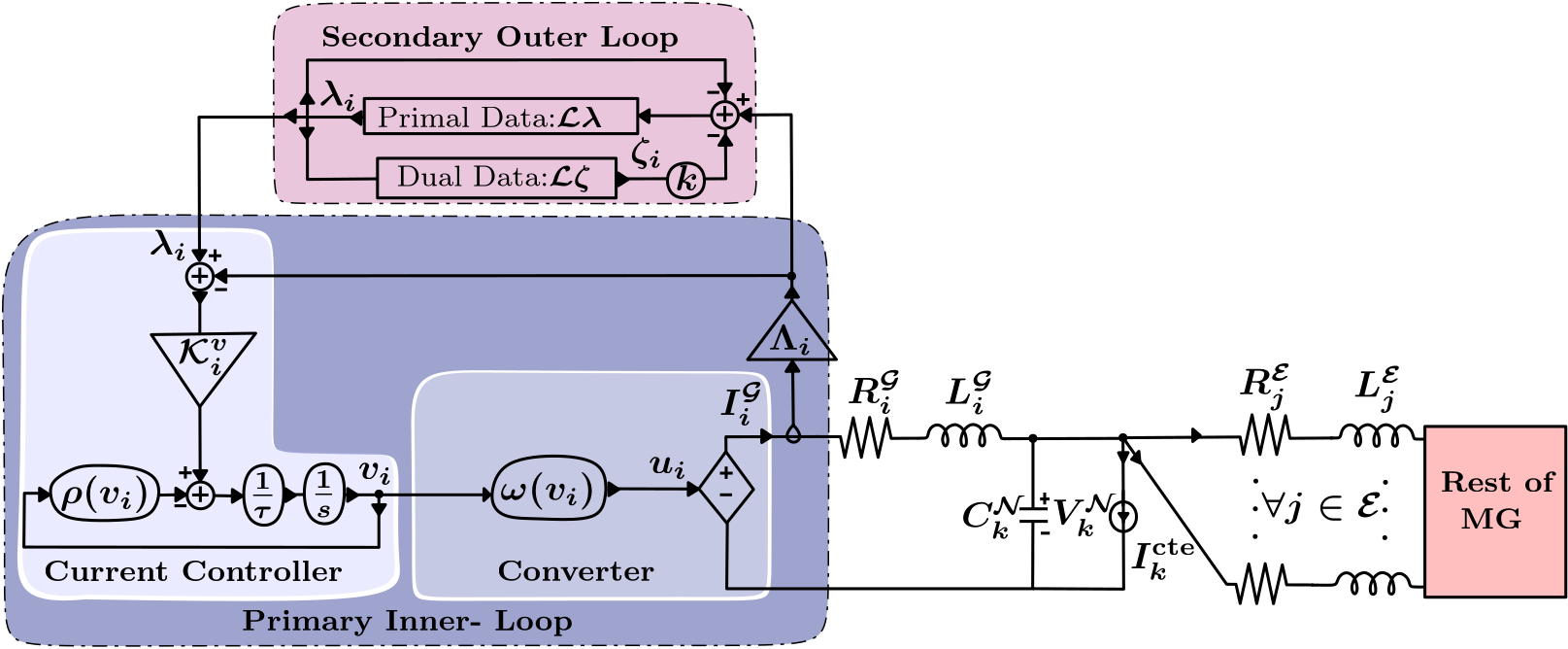}
    \caption{DC microgrid dynamics - A single DG unit perspective.}
    \label{Image:TOT_EL}
\end{figure}
The DC microgrid is based on the models presented in \cite{Babak_Base_Article} and \cite{Babak_Set_Point}. The agents are the distributed generators (DGs), located close to the power-consuming loads (ZI-loads), and are effectively interfaced with the rest of the MG through voltage-controlled converters. Due to the fast dynamics of power electronic converters, their equivalent zero-order models are considered in this paper. The DGs are interconnected both electrically and via distributed communication links, forming a cyber-physical grid. Graph theory is used to establish the physical and virtual interconnections---see Appendix A in \cite{Babak_AC} for the precise definitions of the included graphs. Following the model illustrated in Fig. \ref{Image:TOT_EL}, the electrical dynamics for the $i$-th DG ($i \in \mathcal{G}$), connected to the $j$-th power line ($j \in \mathcal{E}$) and the $k$-th power-consuming load ($k \in \mathcal{N}$) are presented in \eqref{eq:Physical-Layer}.  
\begin{subequations}\label{eq:Physical-Layer}
\begin{align}
L_i^\mathcal{G}\Dot{I}_i^\mathcal{G}&=u_i - \sum_{k} b^\mathcal{G}_{ik}V_k^\mathcal{N}-R_i^\mathcal{G}I_i^\mathcal{G},\\ \label{1b}
L_j^\mathcal{E}\Dot{I}_j^\mathcal{E}&= -\sum_{k} b^\mathcal{E}_{jk}V_k^\mathcal{N}-R_j^\mathcal{E}I_j^\mathcal{E},\\ \label{1c}
C_k^\mathcal{N}\Dot{V}_k^\mathcal{N}&=\sum_{j}b_{kj}^\mathcal{E}I_j^\mathcal{E}+\sum_{i}b_{ki}^\mathcal{G}I_i^\mathcal{G}-G_k^\mathrm{cte}V_k^\mathcal{N}-I_k^\mathrm{cte},
\end{align}
\end{subequations}
where $L_i^\mathcal{G}$, $R_i^\mathcal{G}$, $I_i^\mathcal{G}$, and $u_i$ are the inductance, parasitic resistance, current, and voltage of the \textit{i}th DG, respectively. In \eqref{1b}, $L_j^\mathcal{E}$, $R_j^\mathcal{E}$, and $I_j^\mathcal{E}$ are the inductance, resistance, and current of the \textit{j}th power line, respectively. In \eqref{1c}, $C_k^\mathcal{N}$, $V_k^\mathcal{N}$, $G_k^\mathrm{cte}$, and $I_k^\mathrm{cte}$ respectively are the shunt capacitance, its voltage, the constant conductance, and constant current of the \textit{k}th power-consuming load. The elements $b^\mathcal{G}_{ik}$ and $b^\mathcal{E}_{jk}$ correspond to the incidence matrices, characterizing arbitrary current flows within the network. When current flows from the \textit{j}-th power line to the \textit{k}-th load, the matrix element $b^\mathcal{E}_{jk}$ is set to $1$, in the absence of current exchange $b^\mathcal{E}_{jk}=0$, and 
for an opposite current flow $b^\mathcal{E}_{jk}=-1$. Similarly, the matrix element $b^\mathcal{G}_{ik}$ characterizes current flows between DGs interconnected loads. However, we treat the DGs as power injection components; thus, $b^\mathcal{G}_{ik}=-1$ is not considered.
\subsection{Nested Control Framework}
Motivated by the \emph{nonlinear} controller design in \cite{Babak_Set_Point}, this paper aims to modify the control configuration to guarantee convergence to steady state while ensuring simultaneous achievement of the following two control objectives in DC microgrids:
\begin{itemize}[label=\textendash] 
    \item voltage containment within pre-specified limits,
    \item proportional current sharing at steady state.
\end{itemize}
The voltage containment objective is required to hold uniformly for all admissible initial conditions and disturbances.
These control objectives are mathematically formulated as follows: 
\begin{equation} \label{Control_Obj}
    \begin{split}
        &V_\mathrm{min} \leq u_i(t) \leq V_\mathrm{max}, \quad \forall \; t \in \; \mathbb{R}_{\geq 0},\\
        & \lim_{t \to \infty} \left( I_i^\mathcal{G}(t)/I_i^\mathrm{rated} -I_l^\mathcal{G}(t)/I_l^\mathrm{rated}\right)=0, \quad \forall~i,l \in \mathcal{G}, \ i \ne l,
    \end{split}
\end{equation}
\noindent where $u_i$ is the voltage of the converter of the $i$th DG, as well as its control input, with associated rated current capacity $I_i^\mathrm{rated}$, and $V_\mathrm{min}$ and $V_\mathrm{max}$ are the allowable minimum and maximum values of the voltage output.

To achieve the objectives stated in \eqref{Control_Obj}, a nested control framework is proposed, which comprises two control loops: a decentralized(/primary) inner-loop controller for voltage containment and a distributed(/secondary) outer-loop controller for proportional current sharing.

\subsection{Primary Inner-Loop Controller}
The primary inner-loop controller is designed as a decentralized local current controller, regulating current deviations of the converters with respect to a reference set-point generated by the outer-loop controller. The model incorporates an integral controller, augmented with a \emph{non-permanent} nonlinear leakage function, to ensure consistent system behavior and stable operation, particularly under saturated operation. Additionally, this primary command signal is applied to the converter through a nonlinear hyperbolic tangent function to contain the converter's output voltage. The dynamics of the inner-loop controller for the $i$-th DG are given by:
\begin{subequations} \label{dynamics_inner_loop}
\begin{align}
    \hspace{-.2in}u_i&=\omega(v_i)=V^*+\Delta \mathrm{tanh}(v_i/\Delta), \label{omega}\\ 
    \hspace{-.2in}\tau_i \Dot{v}_i&=-\rho(v_i)v_i+\mathcal{K}^v_i(\lambda_i-\Lambda_iI_i^{\mathcal{G}}), \label{regulator}\\
    \hspace{-.2in}\rho(v_i)&=\alpha (1+0.5[\tanh(\mathrm{b}(v_i-v_\mathrm{pos}))-\tanh(\mathrm{b}(v_i-v_\mathrm{neg}))]), \label{leakage}
\end{align}
\end{subequations}
where $V^*=\frac{1}{2}(V_{\text{min}}+V_{\text{max}})$ is the central point within the safe voltage range, and $\Delta=\frac{1}{2}(V_{\text{max}}-V_{\text{min}})$ is the maximum allowed deviation from $V^*$. The integral controller state $v_i$ enters the control input $u_i$ of DG$_i$ through the nonlinear hyperbolic tangent function $\omega(v_i)$, saturating the controller state value whenever it exceeds predefined limits. In \eqref{regulator}, $\mathcal{K}_i^v>0$ is the integral gain, and $\rho(v_i)$ is a nonlinear leakage operating as an anti-wind-up function when the system reaches voltage saturation under the given arguments; $v_\mathrm{pos}=\Delta\tanh^{-1}[(v_\mathrm{max}-v_\mathrm{tol}-V^*)/\Delta]$ and $v_\mathrm{neg}=\Delta\tanh^{-1}[(v_\mathrm{min}+v_\mathrm{tol}-V^*)/\Delta]$, with $v_\mathrm{tol}$ being the saturation tolerance, $\alpha$ is the leakage coefficient scaling the upper bound value, and $\mathrm{b}$ is implemented to scale the steepness of the curve.
\subsection{Secondary Outer-Loop Controller}
The secondary outer-loop controller adopts a distributed communication-based architecture. Accordingly, by leveraging communication among neighboring DGs, the outer-loop controller restores nominal operating conditions by cooperatively establishing consensus values for the primal and dual communication state variables. Motivated by the control proposal in Section II.C in \cite{Babak_Set_Point}, the outer-loop controller dynamics of the $i$-th DG are given by:
\vspace{-.0in}
\begin{subequations} \label{dynamics_outer_loop}
\begin{align}
    \tau_i^p \Dot{\lambda}_i &= \Lambda_iI_i^{\mathcal{G}}-\lambda_i - \sum_{j\in N^c} [a_{ij} (\zeta_i-\zeta_j)-ka_{ij} (\lambda_j-\lambda_i)], \\
    \tau_i^d \Dot{\zeta}_i &= \sum_{j\in N^c} a_{ij} (\lambda_i-\lambda_j),
\end{align}
\end{subequations}
where $\lambda_i$ and $\zeta_i$ are the primal and dual communicated states of the cyber network with respective time constants $\tau_i^p$ and $\tau_i^d$. In \eqref{dynamics_outer_loop}, $a_{ij}$ is the \emph{adjacency matrix} element, characterizing the communication network topology, $k$ is a positive gain, and $\Lambda_i=1/I_i^\mathrm{rated}$. In summary, this secondary controller is derived by means of the Lagrange multiplier method, which ensures that the system stabilizes at an equilibrium--characterized by the primal and dual Karush–Kuhn–Tucker (KKT) conditions--which is the argument that minimizes a cost function. More precisely, the cost function was designed to ensure proportional current sharing at steady state across the DC MG (see \eqref{Control_Obj}). Consequently, each local optimizer of the outer-loop controller requires the rated current, $\Lambda_i$, of the associated DG$_i$. 

\subsection{Closed-loop DC Microgrid Dynamics}
Given DG dynamics in \eqref{eq:Physical-Layer}, the proposed primary inner-loop controller in \eqref{dynamics_inner_loop}, and the secondary outer-loop controller in \eqref{dynamics_outer_loop}, the closed-loop microgrid dynamics in compact form are presented as follows:
\begin{subequations}\label{complete_sys_compact}
    \begin{align}
    L^\mathcal{G}\Dot{I}^\mathcal{G}&=\omega(v) - \beta^\mathcal{G}V^\mathcal{N}-R^\mathcal{G}I^\mathcal{G} ,\label{complete_sys_compact_I}\\
    L^\mathcal{E}\Dot{I}^\mathcal{E}&= -\beta^\mathcal{E}V^\mathcal{N}-R^\mathcal{E}I^\mathcal{E}, \label{complete_sys_compact_E}\\
    C^\mathcal{N}\Dot{V}^\mathcal{N}&=\beta^{\mathcal{E}\top}I^\mathcal{E}+\beta^{\mathcal{G}\top}I^\mathcal{G}-G^\mathrm{cte}V^\mathcal{N}-I^\mathrm{cte}, \label{complete_sys_compact_Volt} \\
    \tau \Dot{v}&=-\rho(v)v+\mathcal{K}_v(\lambda-\Lambda I^\mathcal{G}), \label{complete_sys_compact_v}\\
    \tau_p \Dot{\lambda}&=\Lambda I^\mathcal{G}-\lambda-\mathcal{L}\zeta-k\mathcal{L}\lambda, \label{complete_sys_compact_lambda}\\
    \tau_d \Dot{\zeta} &= \mathcal{L}\lambda,\label{complete_sys_compact_zeta}
    \end{align}
\end{subequations}
with $L^\mathcal{G}=\mathrm{diag}(L_i^\mathcal{G})\in \mathbb{R}^{n_i\times n_i}$, $I^\mathcal{G}=\mathrm{col}(I_i^\mathcal{G}) \in \mathbb{R}^{n_i}$, $R^\mathcal{G}=\mathrm{diag}(R_i^\mathcal{G}) \in \mathbb{R}^{n_i\times n_i}$, $\omega(v)=\mathrm{col}(\omega(v_i)) \in \mathbb{R}^{n_i}$, $\beta^\mathcal{G}=[b_ki^\mathcal{G}]\in \mathbb{R}^{n_i\times n_k}$. $L^\mathcal{E}=\mathrm{diag}(L_j^\mathcal{E})\in \mathbb{R}^{n_j\times n_j}$, $I^\mathcal{E}=\mathrm{col}(I_j^\mathcal{E}) \in \mathbb{R}^{n_j}$, $R^\mathcal{E}=\mathrm{diag}(R_j^\mathcal{E}) \in \mathbb{R}^{n_j\times n_j}$, $\beta^\mathcal{E}=[b_kj^\mathcal{E}]\in \mathbb{R}^{n_j\times n_k}$. $C^\mathcal{N}=\mathrm{diag}(C_k^\mathcal{N})\in \mathbb{R}^{n_k\times n_k}$, $V^\mathcal{N}=\mathrm{col}(V_k^\mathcal{N}) \in \mathbb{R}^{n_k}$, $G^\mathrm{cte}=\mathrm{diag}(G_k^\mathrm{cte}) \in \mathbb{R}^{n_k\times n_k}$, $I^\mathrm{cte}=\mathrm{col}(I_k^\mathrm{cte}) \in \mathbb{R}^{n_k}$, $v =\mathrm{col}(v^c_{i}) \in \mathbb{R}^{n_i}$, $\tau=\mathrm{diag}(\tau_i) \in \mathbb{R}^{n_i\times n_i}$, $\rho(v)=\mathrm{diag}(\rho(v_i))\in \mathbb{R}^{n_i \times n_i}$, $\mathcal{K}_v=\mathrm{diag}(\mathcal{K}^v_i)\in \mathbb{R}^{n_i\times n_i}$, $\lambda=\mathrm{col}(\lambda_i) \in \mathbb{R}^{n_i}$, $\Lambda=\mathrm{diag}(\Lambda_i) \in \mathbb{R}^{n_i\times n_i}$, $\zeta=\mathrm{col}(\zeta_i) \in \mathbb{R}^{n_i}$, $\tau_p=\mathrm{diag}(\tau^p_i) \in \mathbb{R}^{n_i \times n_i}$, $\tau_d=\mathrm{diag}(\tau_{i}^d) \in \mathbb{R}^{n_i\times n_i}$. Finally, $\mathcal{L}=[l_{ij}]\in \mathbb{R}^{n_i \times n_i}$ is the \emph{Laplacian} matrix, containing the consensus properties. 
\begin{assumption}
\label{ass_Laplacian}\emph{(Laplacian Matrix)}
    It is assumed that the communication network (represented by the Laplacian matrix $\mathcal{L}$) is connected and undirected. 
\end{assumption}

\section{System Steady State and Stability Assessment} \label{Stability}
In this section, we aim to prove that the closed-loop DC MG system augmented with the proposed nested control framework in \eqref{complete_sys_compact} exponentially converges to an equilibrium in which the two control objectives in \eqref{Control_Obj} are simultaneously satisfied. To establish the stability of the closed-loop DC microgrid in \eqref{complete_sys_compact}, while accounting for the nonlinear dynamics and nested control loops, SPT is first employed in Subsection~\ref{seperation}. This verifies the existence of a sufficient time-scale separation between the nested outer- and inner-loop dynamics (including the electrical dynamics), under which a scalable G.E.S. certificate can be derived using Lyapunov-based arguments in Subsection~\ref{stab}.

\subsection{Equilibrium Analysis}\label{section_steady_state} 
To analyze whether the DC microgrid converges to a steady state that satisfies the two control objectives in \eqref{Control_Obj}, we begin by characterizing the equilibrium of \eqref{complete_sys_compact}. To this end, we derive the steady state algebraic equations of \eqref{complete_sys_compact} by imposing $\Dot{x}=0$ for all states. Hence, any steady state needs to satisfy:
\begin{align*}
        \hspace{.5in}0=f(\Bar{I}^\mathcal{G}, \Bar{I}^\mathcal{E}, \Bar{V}^\mathcal{N},\Bar{v}, \Bar{\lambda}, \Bar{\zeta})
\end{align*}
Under \emph{Assumption} \ref{ass_Laplacian}, we exploit the Laplacian property $\mathcal{L}\mathbb{I}=0$. Consequently, the steady state operation of the dual communication dynamics in \eqref{complete_sys_compact_zeta} enforces all DGs to cooperatively reach a consensus on the primal communicated state, i.e., $\Bar{\lambda}=\lambda_s \mathbb{I}$,
where $\lambda_s$ denotes the common consensus value. Considering the steady state operation of \eqref{complete_sys_compact_v} when the voltages operate within the saturation limits; i.e., inactive leakage function: $\rho(\Bar{v})=0$, the normalized DG currents $\Lambda\Bar{I}^\mathcal{G}$ are forced equal to $\bar \lambda$. As a result, the primal communication dynamics in \eqref{complete_sys_compact_lambda} reach steady state when $\Bar{\zeta}=\zeta_s \mathbb{I}$, where $\zeta_s$ denotes the common consensus value, thereby guaranteeing consensus among the DGs with respect to the dual communicated state.
The steady state of the electrical network \eqref{complete_sys_compact_I}-\eqref{complete_sys_compact_Volt} is consequently uniquely defined.
\begin{Remark} \label{lemma_Optimal_steady_state}
\emph{(Optimal Steady State)} Given that the secondary outer-loop controller is designed based on the optimization problem proposed in \cite{Babak_Set_Point}, it follows that the closed-loop DC microgrid converges to an optimal steady state when the primal and dual KKT conditions are satisfied and saturation is avoided. Consequently, $\rho(\bar{v}) = 0$ ensures the ideal proportional current-sharing condition, $\Lambda \bar{I}^{\mathcal{G}} = \bar{\lambda}$, provided that all neighboring DGs collaboratively converge to common consensus values for the primal and dual communication states, and that $\rho(\bar{\lambda})$ is defined uniformly across all DGs. In contrast, during voltage saturation, the non-permanent leakage function in \eqref{complete_sys_compact_v} activates, $\rho(\Bar{v})\neq 0$, which drives the MG to another optimum that satisfies the KKT conditions; however, it no longer achieves ideal proportional current-sharing. Interestingly, the contribution of the leakage function at the converged steady state, $\mathcal{K}_v^{-1}\rho(\bar v)\bar v$, attains the same values as the error between neighboring dual communication variables, $-\mathcal{L}\bar \zeta$, such that the $\Lambda \bar I^\mathcal{G}\neq \bar \lambda$. Note that the voltage containment objective is satisfied uniformly for all operating conditions and admissible initial configurations, enforced by the nonlinear hyperbolic tangent function $\omega(v)$ in \eqref{omega}.
\end{Remark}

\subsection{Time-Scale Separation}\label{seperation}
To facilitate expressing the DC microgrid system as a singularly perturbed problem, the closed-loop system in \eqref{complete_sys_compact} is separated into \emph{slow} and \emph{fast} subsystems under the time-scale separation assumption stated below:
\begin{assumption}\emph{(Time-Scale Separation)}
\label{Asump_time_Const} 
It is assumed that the outer-loop dynamics operate at a faster time-scale than the rest of the system dynamics. The outer-loop's respective time constants ($\tau_p, \; \tau_d$) are therefore assumed to be significantly smaller than the smallest time constant of the slow dynamics ($\varepsilon$). Thus, we assume: $\tau_p, \tau_d \ll \varepsilon$, with $\varepsilon$ assumed to be the time constant of the integral controller, $\tau$. 
\end{assumption}
Specifically, under {\it Assumption}~\ref{Asump_time_Const}, the DC microgrid dynamics in \eqref{complete_sys_compact_I}–\eqref{complete_sys_compact_v} constitute the slow subsystem $s(\cdot)$, whereas \eqref{complete_sys_compact_lambda}–\eqref{complete_sys_compact_zeta} represent the fast dynamics$f(\cdot)$--forming the two time-separated systems:
\begin{align} \label{two_separated}
        \hspace{.45in}\Dot{x}=s(x, z) \quad \text{and} \quad 
        \varepsilon \Dot{z}=f(x, z),
\end{align}
with $x=\mathrm{bcol}\{I^\mathcal{G}, I^\mathcal{E}, V^\mathcal{N}, v\}\in \mathbb{R}^{2 n_i+n_j+n_k}$ and $z=\mathrm{bcol}\{\lambda, \zeta\} \in \mathcal{R}^{2 n_i}$.

To further simplify the stability analysis, the order of the outer-loop is preliminarily reduced using a method similar to \cite{Babak_AC}. The specific assumptions underlying this order reduction are detailed in the following.
\begin{assumption}\emph{(Time-scales of the Outer-Loop)}
\label{Asump_outer_Loop}  The primal communication dynamics in \eqref{complete_sys_compact_lambda} are assumed to operate at a significantly faster time-scale than the dual communication dynamics in \eqref{complete_sys_compact_zeta}--characterized by the following time-scale assumption: $\tau_p \ll \tau_d$.
\end{assumption}
\vspace{-.07in}
Let \emph{Assumption} \ref{Asump_outer_Loop} hold, and assume that $\tau_p \Dot{\lambda}\approx 0$. As a result, the primal communication state can be expressed as a function of the dual communication state, i.e.,  $\lambda=K_1\Lambda I^\mathcal{G}-K_1\mathcal{L}\zeta$, with a positive-definite matrix $K_1\triangleq[\mathbb{I}+k\mathcal{L}]^{-1} > 0$, resulting in some added damping in the $\zeta$ coordinate in \eqref{complete_sys_compact_zeta}. Accordingly, the closed-loop DC MG dynamics--specifically \eqref{complete_sys_compact_v}–\eqref{complete_sys_compact_zeta}--are redefined as follows:
\begin{subequations}
\begin{align}
    \hspace{-.2in}\tau \Dot{v}&=-\rho(v)v+\mathcal{K}_v(K_1(\Lambda I^\mathcal{G}-\mathcal{L}\zeta)-\Lambda I^\mathcal{G})-\mathcal{B}_v(v-v^*),\label{reduced_complete_d}\\
    \hspace{-.2in}\tau_d\Dot{\zeta} &= \mathcal{L}K_1(\Lambda I^\mathcal{G}-\mathcal{L}\zeta).\label{reduced_complete_e}
\end{align}
\end{subequations}
Consequently, the two time-separated systems in \eqref{two_separated} are now governed by the slow states $x$ and the \emph{reduced} fast state vector $z^*=\mathrm{col}(\zeta_i), \ \forall i \in \mathbb{R}^{n_i}$:
\begin{align} \label{two_separated_reduces}
        \hspace{.3in}\Dot{x}=s(x, z^*) \quad \text{and} \quad 
        \varepsilon \Dot{z}^*=f(x, z^*).
\end{align}
Additionally, a \emph{permanent} leakage term $\mathcal{B}_v=\mathrm{diag}(\mathcal{B}^v_i)\in \mathbb{R}^{n_i\times n_i}$ is introduced in the integral controller state coordinate in \eqref{reduced_complete_d} to introduce dissipation when the voltages operates within safe limits ($\rho(v)=0$), which is essential for the subsequent stability analysis, with $v^*$ being the constant estimated value of $v$ in steady state for the ideal nominal operating point. 
\begin{assumption}\emph{(Frozen Variables in the Boundary Layer System)} \label{ass:frozen}
Let Assumptions \ref{Asump_time_Const} and \ref{Asump_outer_Loop} hold, and let 
\begin{align*}
    \mathrm{h}(x)=\mathrm{bcol}\{\mathrm{h}_i(x)\}, \ \forall i \in \mathcal{N}_c,
\end{align*}
be the unique solution of $\varepsilon \dot z^*=f(x, z^*)$ for $\varepsilon \approx 0$.
Expressed in terms of incremental states, it then gives
    \begin{align*}
        \hspace{.18in}\mathrm{H}(\tilde x)&\triangleq
        \mathrm{bcol}\{ \mathrm{h}(x)-\mathrm{h}(\bar x)\}= \mathrm{bcol}\{\hat{\mathrm{H}}(\tilde x)-\mathrm{h}(\bar x)\},
    \end{align*}
    with $\tilde{x}=x - \bar x$, and $\hat{\mathrm{H}}(\tilde x)=\mathrm{h}(\tilde x+\bar x)$. Under Assumption~\ref{Asump_time_Const}, we assume that the slow states entering the fast dynamics can be treated as frozen variables. Thus, in the boundary layer system, $x=\bar x$ and therefore $\tilde x = 0$, resulting in 
    $$\hat{\mathrm{H}}(\tilde x)=\mathrm{h}(\bar x) 
    \rightarrow \mathrm{H}(\tilde x)=\{\mathrm{h}(\bar x)-\mathrm{h}(\bar x)\}=0.$$
    Moreover, let $\tilde y\triangleq \mathrm{bcol}\{\tilde z^*- \mathrm{H}(\tilde x)\}$ be the error between the actual fast dynamics $(\tilde z^*)$ and the quasi-steady state $(\mathrm{h}(\tilde x))$. Under sufficient time-scale separation, we assume that $\bar z^*= \mathrm{h}(\bar x)$ such that $\bar y=0$ in the boundary layer system. Thus, from the above expressions, we let 
    \begin{align*}
        \hspace{.15in}\Tilde{\mathbf{z}}^*= (y-\bar y)+(\mathrm{h}(x)-\mathrm{h}(\bar x))=\tilde y, \quad \text{with} \quad y=\tilde y + \bar y.
    \end{align*}
\end{assumption} 
\begin{Theorem}\emph{(Singularly Perturbed Problem)}\label{theorem_SPT} Let Assumptions \ref{ass_Laplacian}-\ref{ass:frozen} hold. Consider the closed-loop dynamics of the reduced model in \eqref{two_separated_reduces}, and let 
    \begin{equation} \label{theorem_def}
        \begin{split}
        &\Omega(\tilde v)  \triangleq \omega(v)-\omega(\Bar{v}), \hspace{0.4in} \Gamma(\Tilde{v})\triangleq\rho(v)v-\rho(\Bar{v})\Bar{v},\\
        &\tilde{\hat x}\triangleq \mathrm{bcol}\{\tilde{I}^\mathcal{G}, \tilde{I}^\mathcal{E},  \tilde{V}^\mathcal{N} \}, \hspace{0.3in} 
        \mathcal{P}_s \triangleq \mathrm{bdiag}\{R^\mathcal{G},R^\mathcal{E},G^\mathrm{cte} \}> 0,\\
        &\mathcal{Q}_f\triangleq\mathrm{diag}(\tau_i^d)> 0, \hspace{0.41in} \mathcal{P}_f\triangleq\mathcal{L}K_1\mathcal{L}> 0,
        \\
        &\mathcal{Q}_s \triangleq\mathrm{bdiag}\{L_i^\mathcal{G},L_j^\mathcal{E},C_k^\mathcal{N} \}> 0,\\
        &\mathcal{J}_s\triangleq\begin{bmatrix}
        0 & 0 & -\beta^\mathcal{G}\\
        0 & 0 & -\beta^\mathcal{E}\\
        \beta^{\mathcal{G}\top} & \beta^{\mathcal{E}\top} & 0 \end{bmatrix}, \hspace{0.1in}  \kappa_1 \triangleq \mathrm{bcol}\{\mathbb{I},0,0\}.
        \end{split}
    \end{equation}
The two time-separated systems in \eqref{two_separated_reduces} can then be expressed as the singular perturbed problem in \eqref{SP_Compact_Inc} under the stretched timescale $\mathbf{t}=(t/\varepsilon)$: 
\begin{subequations}\label{SP_Compact_Inc}
\begin{align}
    \hspace{-0.3in}\hat s(\tilde x, \mathrm{H}(\tilde x))&:
                \begin{cases}
                \mathcal{Q}_s \dot{\tilde{ \hat x}}=(\mathcal{J}_s-\mathcal{P}_s)\Tilde{\hat x} + \kappa_1 \Omega(\tilde v),\\
                \tau \dot{\tilde v}=-\Gamma(\tilde v)+\mathcal{K}_v(K_1(\Lambda \tilde{I}^\mathcal{G}-\mathcal{L}\mathrm{H}(\tilde x))-\Lambda\tilde{I}^\mathcal{G})-\mathcal{B}_v\tilde v, 
                \end{cases} \label{RS_compact_inc}\\
    \hspace{-0.3in}\hat f(\tilde y) &: 
                \begin{cases}
                \mathcal{Q}_f \partial \Tilde{y}/\partial \mathbf{t} = -\mathcal{P}_f\Tilde{y}, \label{BL_compact_inc}
                 \end{cases}
\end{align}
\end{subequations}
where $\hat{s}(\tilde x, \mathrm{H}(\tilde x))$ and $\hat{f}(\tilde y)$ represent the reduced and boundary layer systems, respectively.
\end{Theorem}
\begin{proof}
     First, we consider $\varepsilon$ to be significantly small such that the velocity of $\Dot{z}\propto (1/\varepsilon)$ behaves instantaneously fast. Thus, for $\varepsilon\approx 0$, the fast dynamics in \eqref{reduced_complete_e} quickly reaches a \emph{quasi-steady state} given the instantaneous fast dynamics $\mathrm{h}(x)=\mathrm{col}(\mathrm{h}_i(x)), \forall i\in \mathbb{R}^n_i$. We further define $y=\mathrm{col}(y_i), \forall i\in \mathbb{R}^n_i$ as the error between the \emph{actual} fast dynamics and the quasi-steady state: $y\triangleq \mathrm{col} (z^*-\mathrm{h}(x))$. Moreover, to facilitate later application of Lyapunov theory, we express the system states using their incremental variables $\tilde z^*= z^* - \bar z^*$, $\tilde y=y-\Bar{y}$ and $\tilde x=x-\Bar{x}$. Therefore, when expressed in terms of incremental states, the \emph{actual} fast dynamics are given by $\Tilde{\mathbf{z}}^* \triangleq \tilde y + \mathrm{h}(\tilde x + \bar x) - \mathrm{h}(\bar x)=\tilde y + \mathrm{H}(\tilde x)$, and we express the two time-separated systems in \eqref{two_separated_reduces} as: 
    \begin{align} \label{slow/Fast_time_sep}
        \hspace{-.25in}\Dot{\tilde{x}}=s(\tilde x, \tilde y + \mathrm{H}(\tilde x)), \quad \text{and} \quad 
        \varepsilon \frac{\partial (\tilde y + \mathrm{H}(\tilde x))}{\partial t}=f(\tilde x, \tilde y + \mathrm{H}(\tilde x)).
    \end{align}
    Under the stretched time-scale $\mathbf{t}=(t/\varepsilon)$--where $t$ is the time when $\varepsilon \approx 0$--the two time-separated systems in \eqref{slow/Fast_time_sep} can be written as the \emph{singular perturbation problem} in \eqref{SP_Compact_Inc}, divided into the reduced system \eqref{RS_compact_inc} and the boundary layer system \eqref{BL_compact_inc}. Under this stretched time-scale, the slow dynamics instantaneously attains a quasi–steady state as the fast dynamics are considered instantaneously fast, implying that $\tilde y\approx 0$ in the reduced system. Furthermore, if \emph{Assumption}~\ref{ass:frozen} holds, the slow states entering the boundary layer are considered frozen variables, implying that $\tilde x=0$, $\bar y=0$, such that $\mathrm{H}(\tilde x)=0$, and $\Lambda \tilde{I}^\mathcal{G}=0$, in the boundary layer system \eqref{BL_compact_inc}. 
\end{proof}

\subsection{Stability Analysis}\label{stab}
The following theorem analyzes the stability of the closed-loop DC microgrid in \eqref{complete_sys_compact}, presented as a singular perturbed system in \eqref{SP_Compact_Inc}.
\begin{assumption}(\emph{Monotonicity}) \label{Ass_mon}
The nonlinear function $\Omega(\tilde{v})$ is assumed to be strictly monotonically increasing, while $\Gamma(\tilde{v})$ is assumed to be, at the very least, monotonically increasing.
\end{assumption}
\begin{Theorem} \label{theorem_GES}\emph{(Exponential Stability of the Singularly Perturbed system).} Consider the singular perturbed system in \eqref{SP_Compact_Inc} and suppose that $\Omega(\tilde v)$ and $\Gamma (\tilde v)$ are monotonically increasing in $\tilde v$. Let
\begin{equation}\label{eq13}
    \begin{split}
&K_1\triangleq[\mathbb{I}+k\mathcal{L}]^{-1}, \hspace{0.63in}
    K_2\triangleq [\mathcal{L}K_1\mathcal{L}]^{-1}\mathcal{L}K_1,  \\
    &K_4\triangleq-\frac{1}{2}[\mathbb{I}+\Lambda K_3^\top],  \hspace{0.43in} K_3\triangleq [\mathcal{K}_v(K_1-K_1\mathcal{L}K_2-\mathbb{I})],\\
    &\mathcal{M}_x(x)\triangleq\begin{bmatrix} \mathcal{P}_x\hat x - \kappa_1[\mathbb{I}+\Lambda K_3^\top]\omega(v)\\  \mathcal{B}_v\omega({v})
    \end{bmatrix}, \hspace{0.3in}  x\triangleq \begin{bmatrix} \hat x\\v\end{bmatrix},
\end{split}
\end{equation}
and $\mathcal{B}^*$ be the maximum eigenvalue of $K_4^\top{R}^{\mathcal{G}-1}K_4 \omega'_\mathrm{max}$ for $\omega'_\mathrm{max}=\mathrm{max}(\partial \omega(v)/\partial v)$.
If $\mathcal{M}_x(x)$ is strictly monotonically increasing and the reduced system in \eqref{two_separated_reduces} satisfies the following matrix inequality
\begin{align}
    \hspace{.5in}\mathcal{B}_v \geq \mathcal{B}_v^* \geq K_4^\top R^{\mathcal{G}^{-1}} K_4 \omega'(v), \label{stab_condition}
\end{align}
then, there exists $\varepsilon^*>0$ such that for all $\tau_d\ll\varepsilon<\varepsilon^*$ the system in \eqref{two_separated_reduces} is globally exponentially stable under the following bound: 
\begin{align}
    \hspace{-.2in}-(x-\bar x)^\top \left[\mathcal{M}_x({x}) - \mathcal{M}_x({\bar x})\right]\leq -\Upsilon ||\Tilde{x}||^2, \quad \Upsilon \; \in \mathbb{R}_{>0}.\label{Bound_2}
\end{align}
\end{Theorem}
\begin{proof}
    From the dynamics of the singularly perturbed system in \eqref{SP_Compact_Inc}, we propose the following Lyapunov candidates for the reduced system; $V_s$, and for the boundary layer system; $W_f$.  
    \begin{subequations}
    \begin{align}
    &\hspace{.3in}V_s(\Tilde{x})=\frac{1}{2}\Tilde{\hat x}^\top \mathcal{Q}_s\Tilde{\hat x} + \int_{0}^{\Tilde{v}}\Omega^\top(\sigma) \tau d\sigma \geq 0,\label{lyap_slow}\\
       &\hspace{.6in}W_f(\Tilde{y})=\frac{1}{2}\Tilde{y}^\top\mathcal{Q}_f\Tilde{y}\geq 0,\label{lyap_fast}
    \end{align}
    \end{subequations}
    where $\mathcal{Q}_s$, $\mathcal{Q}_f$, and $\Omega(.)$ are defined in \eqref{theorem_def}. As $\mathcal{Q}_f$ is a positive-definite matrix, $W_f>0$, $\forall \tilde y \neq 0$, and is radially unbounded. Moreover, $\mathcal{Q}_s > 0$ ensures that the first quadratic term of \eqref{lyap_slow} is positive. Examining the sign of the integral part in \eqref{lyap_slow}, $\Omega(\tilde v)$ is defined in \emph{Theorem}~\ref{theorem_SPT} as a strictly monotonically increasing function in $\tilde v$. Accordingly, when multiplied by the increment of the curve ($\sigma$), the integral is guaranteed to be positive for all $\tilde v$. As a result, $V_s$ is positive for all $\forall \tilde v\neq 0$, and radially unbounded.
    
    The time derivative of the Lyapunov function associated with the reduced system is then obtained as follows:
    \begin{align}\label{deriv_Lyap_slow}
        \Dot{V}_s(\Tilde{x})&=\nabla^\top V_s(\Tilde{x})\Dot{\Tilde{x}}\nonumber\\
        &=-\Tilde{\hat x}^\top \mathcal{P}_s  \Tilde{\hat x} + \Tilde{\hat x}^\top \kappa_1 \Omega(\tilde v) - \Omega^\top(\tilde v)\Gamma(\tilde v) \nonumber\\
        &-\Omega^\top (\tilde v)\mathcal{B}_v \tilde v + \Omega^\top (\tilde v) K_3 \Lambda \kappa_1^\top \Tilde{\hat x} \nonumber\\
        &= -(x-\bar x)^\top\left[ \mathcal{M}_x({x})-\mathcal{M}_x({\bar x})\right] - \Omega(\Tilde{v})^\top \Gamma(\Tilde{v})\nonumber \\
        &\leq - (x-\bar x)^\top\left[ \mathcal{M}_x({x})-\mathcal{M}_x({\bar x})\right]\leq 0.
    \end{align}
    In the above derivation, the structure of the reduced system has been modified so that the Lyapunov function is solely a function of the slow states. Specifically, the integral controller dynamics in \eqref{RS_compact_inc} depend on the instantaneous fast dynamics $\mathrm{H}(\tilde x)$. However, solving $\varepsilon \Dot{\tilde z}^*=f(\tilde x, \tilde z^*)$ for $\varepsilon\approx 0$ yields a closed-form expression of the instantaneous fast dynamics given as a function of the slow state $\tilde I^\mathcal{G}$. Hence, $\mathrm{H}(\tilde x)=K_2\Lambda \tilde I^\mathcal{G}$, allowing the dynamics of the integral controller to be expressed as: 
    \begin{align*}
        \hspace{.5in}\tau \Dot{\tilde v}=-\Gamma(\tilde v)+K_3\Lambda \tilde I^\mathcal{G}-\mathcal{B}_v\tilde v,
    \end{align*}
    with $K_2$ and $K_3$ defined in \eqref{eq13}. Moreover, $\mathcal{P}_s$ is a positive-definite matrix (see \eqref{theorem_def}) and the first quadratic term in the second equality in \eqref{deriv_Lyap_slow} is non-positive. Furthermore, $\Omega(\Tilde{v})$ and $\Gamma(\Tilde{v})$ are monotonically increasing in $\Tilde{v}$, ensuring that their product is positive such that $- \Omega(\Tilde{v})^\top \Gamma(\Tilde{v})\leq 0$\footnote{Note that the monotonicity \emph{Assumption}~\ref{Ass_mon} is always  (strictly) verified for $\Omega(\tilde v)$ since $\Omega'(\tilde v)>0$, while \emph{practically} verified for $\Gamma(\tilde v)$ for \emph{certain tuning} choices of $b$, $v_\mathrm{pos}$, and $v_\mathrm{neg}$ by; (i) ensuring a pronounced steepness of $\rho(v)$ via $b$, (ii) sufficient separation between $v_\mathrm{pos}$ and $v_\mathrm{neg}$. }. Consequently, $\Dot{V}_s(\Tilde{x}) \le 0$ and the Lyapunov function is guaranteed to be decreasing along the trajectories of the reduced system if and only if $\mathcal{M}_x(x)$ is monotonically increasing, that is, if
    \begin{align*}
        \hspace{-.2in}\tilde x^\top\left(\mathcal{M}_x(x)-\mathcal{M}_x(\bar x) \right)\geq0 \Leftrightarrow \frac{1}{2}\left[\frac{\partial\mathcal{M}_x(x)}{\partial x} + \frac{\partial\mathcal{M}_x(x)}{\partial x}^\top\right] > 0.
    \end{align*}
     Computing the symmetric part of $\frac{\partial\mathcal{M}_x(x)}{\partial x}$ then gives
     \begin{align} \label{Mx}
        \left.\frac{\partial \mathcal{M}_x}{\partial x}\right|_\text{sym} = \begin{bmatrix}
            R^\mathcal{G} & 0 & 0 & K_4 \omega'(v)\\
            0 & R^\mathcal{E} & 0 & 0 \\
            0 & 0 & G^\mathrm{cte} & 0 \\
            \omega'(v)K_4^\top & 0 & 0 & \mathcal{B}_v\omega'(v)
        \end{bmatrix},
    \end{align}
    \noindent with $\omega'(v)=\partial \omega(v)/\partial v$ and $K_4$ defined in \eqref{eq13}. Applying the Schur Complement to \eqref{Mx} provides the following inequality conditions for the matrix to be positive-definite: 
    \begin{align*}
        \hspace{.2in}&\mathcal{B}_vw'(v)-\omega'(v)H^\top R^{\mathcal{G}^{-1}}K_4\omega'(v) > 0, \\
        \hspace{.2in}&\hspace{0.4in}\mathcal{B}_v >  K_4^\top R^{\mathcal{G}^{-1}}K_4\omega'(v),
    \end{align*}
    where the last inequality follows from multiplying the previous condition by the inverse of $\omega'(v)$. 
 Given the sigmoid nature of $\omega(v)$ in \eqref{omega}, the nonlinear function is bounded by $-1\leq \tanh(\cdot) \leq 1$, such that  $0<\omega'(v)\leq \omega'_\mathrm{max}$, which implies that $\mathcal{B}_v$ can be chosen to be greater than a \emph{constant} value $\mathcal{B}_v^*$, defined as the largest eigenvalue of $K_4^\top R^{\mathcal{G}^{-1}}K_4\omega'_\mathrm{max}$, with $\omega'_\mathrm{max}=\mathbb{I}$, which is indeed the condition in \emph{Theorem} \ref{theorem_GES}. Accordingly, we conclude upon global asymptotic stability, provided that the stability condition in \eqref{stab_condition} is satisfied. Furthermore, to guarantee G.E.S., we need to find an upper bound of the Lyapunov function $V_s(\tilde x) \leq c||\tilde x||^2$.
 Exploiting the fact that $\Omega(\tilde v)$ adheres to the Lipschitz condition, we contain the value of the integral as follows
\begin{align}
    \hspace{.5in}\int_{0}^{\tilde v}[\Omega(\sigma)]^\top \partial \sigma \leq \mathsf{L} \int_{0}^{\tilde v} \sigma  \;\; \partial \sigma, \nonumber
\end{align}
for some scalar $\sigma > 0$ and $\mathsf{L}=1$ is the Lipschitz constant. Thus, $V_s(\tilde x) \leq V_1(\tilde x)$ with $V_1(\tilde x) \triangleq \frac{1}{2}\Tilde{\hat x}^\top Q_s \Tilde{\hat x} + \frac{1}{2} \mathsf{L} \tilde v$ $= \frac{1}{2}\tilde x^\top \mathcal{Q}_s^*\tilde x\leq \Phi||\tilde x ||^2$ and  $\mathcal{Q}_s^*\triangleq \mathrm{bdiag}\{\mathcal{Q}_s, \mathsf{L}\}>0$, with $\Phi$ being the maximum eigenvalue of $\mathcal{Q}_s^*$. Hence, we have 
\begin{align} \label{X1}
   \hspace{.8in} V_s(\tilde x)\leq \Phi ||\tilde x||^2.
\end{align}
Finally, using \eqref{X1} and \eqref{Bound_2}, we conclude upon G.E.S. of the reduced  system with a convergence rate given by $\epsilon=2 \frac{\Upsilon}{\Phi}.$
Considering the Lyapunov function of the boundary layer system--given in \eqref{lyap_fast}--its time derivative is given as:
\begin{align}
    \hspace{.2in} \frac{\partial W_f(\tilde y)}{\partial \mathbf{t}}=\nabla^\top W_f(\Tilde{y})\frac{\partial \Tilde{y}}{\partial \mathbf{t}}=- \Tilde{y}^\top \mathcal{P}_f\Tilde{y}\leq 0,\label{derive_Lyap_fast}
\end{align}
with $\mathcal{P}_f$ defined to be positive definite in \emph{Theorem}~\ref{theorem_SPT}, ensuring that the final inequality in \eqref{derive_Lyap_fast} holds. Accordingly, the singularly perturbed system in \eqref{SP_Compact_Inc} satisfies all conditions given in \emph{Theorem} 11.4 in \cite{Khalil} and \emph{Theorem}~\ref{theorem_GES}, and the system in \eqref{complete_sys_compact} is globally exponentially stable as there exists a $\varepsilon^*>0$ for all $\tau_p\ll\tau_d\ll\varepsilon<\varepsilon^*$ 
\end{proof}

\subsection{Practical Guidelines for Stability Guarantees}\label{sec:beta} 
The previous section establishes sufficient \emph{structurally scalable} stability conditions under which the steady state operation of the closed-loop $n$-generator DC microgrid is globally exponentially stable. Specifically, the resulting G.E.S. certificate remains valid for any scale of the system provided that (i) sufficient time-scale separation is imposed between the nested control loops, as specified in \emph{Assumption}~\ref{Asump_time_Const}, and (ii) the integral controller incorporates a \emph{permanent} leakage term whose value satisfies the stability condition in \eqref{stab_condition}. 

\begin{figure}
    \centering
    \includegraphics[width=\columnwidth]{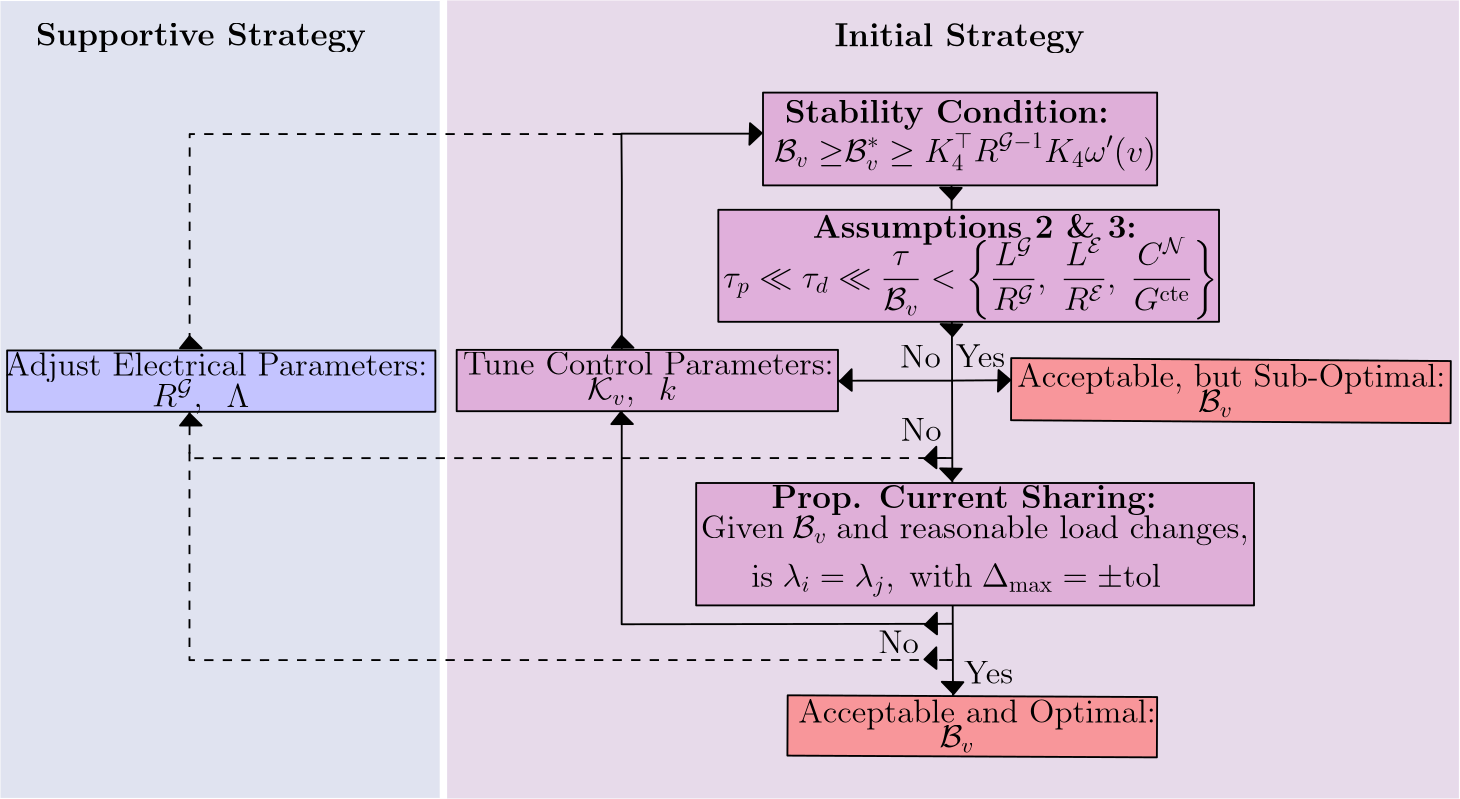}
    \captionsetup{font=small}
    \caption{Practical guidelines for stability and proportional current sharing.}
    \label{Fig-TA}
\end{figure}

Accordingly, from a practical perspective, this subsection proposes two \emph{practical} strategies--illustrated in Fig.~\ref{Fig-TA}--to carefully establish sufficient \emph{system configurations} under which (i) the stability condition in \eqref{stab_condition} of Theorem~\ref{theorem_GES} holds $\forall t \in \mathbb{R}_{\geq 0}$ while enforcing \emph{practically scalable} properties, (ii) sufficient time-scale separation is preserved, and (iii) the selected $\mathcal{B}_v$ is minimized to ensure that, at the converged equilibrium the current integration error in \eqref{complete_sys_compact_v} deviates by no more than $\pm \mathrm{tol}$ from the \emph{optimal} equilibrium where \emph{ideal} proportional current sharing is satisfied--further referred to as \emph{practical} proportional current sharing. However, it should be noted that a trade-off exists between satisfying these three objectives. In particular, following \emph{Remark}~\ref{lemma_Optimal_steady_state}, any active leakage term introduced in the integral controller dynamics in \eqref{complete_sys_compact_v} inhibits \emph{ideal} proportional current sharing at steady state. Moreover, per  \emph{Assumption}~\ref{ass:tau} and \emph{Remark}~\ref{rem_time_scale}, the effective time constant of the integral controller is modified to $\tau_i/\mathcal{B}_i^v$. Consequently, an increased value of $\mathcal{B}_i^v$, required to satisfy \eqref{stab_condition}, may both degrade \emph{practical} proportional current sharing and violate the required time-scale separation. Accordingly, the two proposed strategies provide a means of balancing these competing objectives, depending on the system operator's priorities.
\begin{assumption}\emph{(Approximated Time Constant})\label{ass:tau}
    The proposed evaluation of the time-scales relies on the approximation where the convergence rate of each state can be defined by the ratio of inertia to damping $\tau_i=\mathcal{I}_i/\mathcal{D}_i$, provided that sufficient damping is introduced in the associated system dynamics.
\end{assumption}
\begin{Remark} \label{rem_time_scale} \emph{(Modified Time Constants)}
    When the state under consideration exhibits relatively strong damping, we let \emph{Assumption} \ref{ass:tau} hold. However, when the dynamical state exhibits negligible or \emph{non-permanent} damping, the time constant is defined by the parameter multiplying the time derivative of the state. 
\end{Remark}
\subsubsection{Initial Strategy}
From the stability condition in \eqref{stab_condition}, it is shown that the value of the elements of $\mathcal{B}_v$ ($\mathcal{B}^v_i$) is determined by the $i$th generators resistance ($R_i^\mathcal{G}$), rated current ($\Lambda_i$), the integral gain ($\mathcal{K}_i^v$), and the positive control gain ($k$). Accordingly, to avoid dependence on specific system parameters and improve the \emph{practical scalability} and generic applicability of the results presented in this paper, the initial (and preferred) strategy in Fig.~\ref{Fig-TA} relies solely on careful controller tuning to reduce $\mathcal{B}_i^v$. Among the available tuning parameters, reducing $\mathcal{K}_i^v$ has the greatest effect on decreasing $\mathcal{B}^v_i$, thereby providing the most effective means of achieving a satisfactory reduction. However, when analyzing the steady state operation of the outer-loop controller in \eqref{dynamics_outer_loop}, it becomes evident that maintaining a sufficiently large value of $\mathcal{K}_i^v$ reduces the contribution of the \emph{permanent} leakage at the converged equilibrium, thereby minimizing its adverse impact on the proportional current sharing objective (see \emph{Remark}~\ref{lemma_Optimal_steady_state}). Consequently, from a performance perspective, an acceptable value for $\mathcal{B}_i^v$ may be \emph{sub-optimal} if the system converges to a steady state where the current integration errors deviate beyond the specified \emph{practical} tolerance.

\subsubsection{Supportive Strategy}
If the initial strategy is insufficient in obtaining a controller tuning satisfying the stability conditions (or the \emph{practical} proportional current sharing objective), there exists a supportive strategy that includes modifying the associated system parameters ($R^\mathcal{G}$, $\Lambda$), albeit at the cost of generating a more system-dependent stability certificate. Alternatively, as stated by \emph{Remark}~\ref{rem_Virtual_Impedance}, the supportive strategy may be generalized such that the G.E.S. certificate remains valid without modifying specific system parameters.
\begin{Remark}\label{rem_Virtual_Impedance} \emph{(Virtual Impedance Loop)}
    Whenever the initial controller tuning strategy is insufficient to guarantee the stability condition in \eqref{stab_condition} (or practically guarantee proportional current sharing), a virtual impedance loop can be included in the primary inner-loop controller in \eqref{omega} as an alternative to the supporting strategy. In this case, a digitally implemented virtual impedance can effectively act like an increased generator resistance, thereby reducing the value of the elements of $\mathcal{B}_v$ without requiring specific system parameters. 
\end{Remark}
\begin{Remark}\emph{(Practically Scalable System Configurations)}
Given that the values of $\mathcal{B}_v$ are determined by the rated currents of the DGs and their parasitic resistances, the selected system configuration may appear case-specific, thereby potentially limiting scalability in practice. However, the proposed practical guidelines aim to select the minimum value of $\mathcal{B}_v \geq \mathcal{B}_v^*$ that guarantees stability while preserving practical proportional current sharing. Hence, if $\mathcal{B}_v=\mathcal{B}_v^*$ and the corresponding value satisfies the time-scale separation requirement, stability is guaranteed for the identified worst-case operating condition. Consequently, stable plug-and-play operation is ensured for system expansions involving DGs with similar power ratings; i.e., the system is \emph{practically scalable} within a certain range of operating conditions and parameter settings. Moreover, increasing the rated current capacities and parasitic resistances of the DGs reduces the minimum required value of the elements of $\mathcal{B}_v$. Therefore, the DC microgrid remains \emph{practically scalable} for sufficiently broad system configurations, including expansions with additional DGs (and loads) having similar or higher rated capacities than those considered in the case-specific DC microgrids investigated in the subsequent case studies.
\end{Remark}
\begin{Remark}\emph{(Communication Rate Requirements)}
    It is important to note that, given the already high-speed nature of DC systems, very fast communication rates are essential to uphold this time-scale separation. Interestingly, recent advancements in electric intelligent vehicular technology have led to the development of more reliable and faster communication protocols to enhance the safety and performance of autonomous vehicles. To deal with this growing demand for data transmission, the Controller Area Network with Flexible Data Rate (CAN-FD) communication protocol has been introduced, supporting data rates of up to 8 Mbps \cite{Com1, Com2, Com3}. Although potentially limited to some (geographically-contained) applications, the CAN-FD protocol provides sufficient communication rates supporting the proposed time-scales proposed in this paper. 
\end{Remark}
\section{Case Studies} \label{sec_Case_Studies}
\begin{figure}
    \centering
    \includegraphics[width=.93\columnwidth]{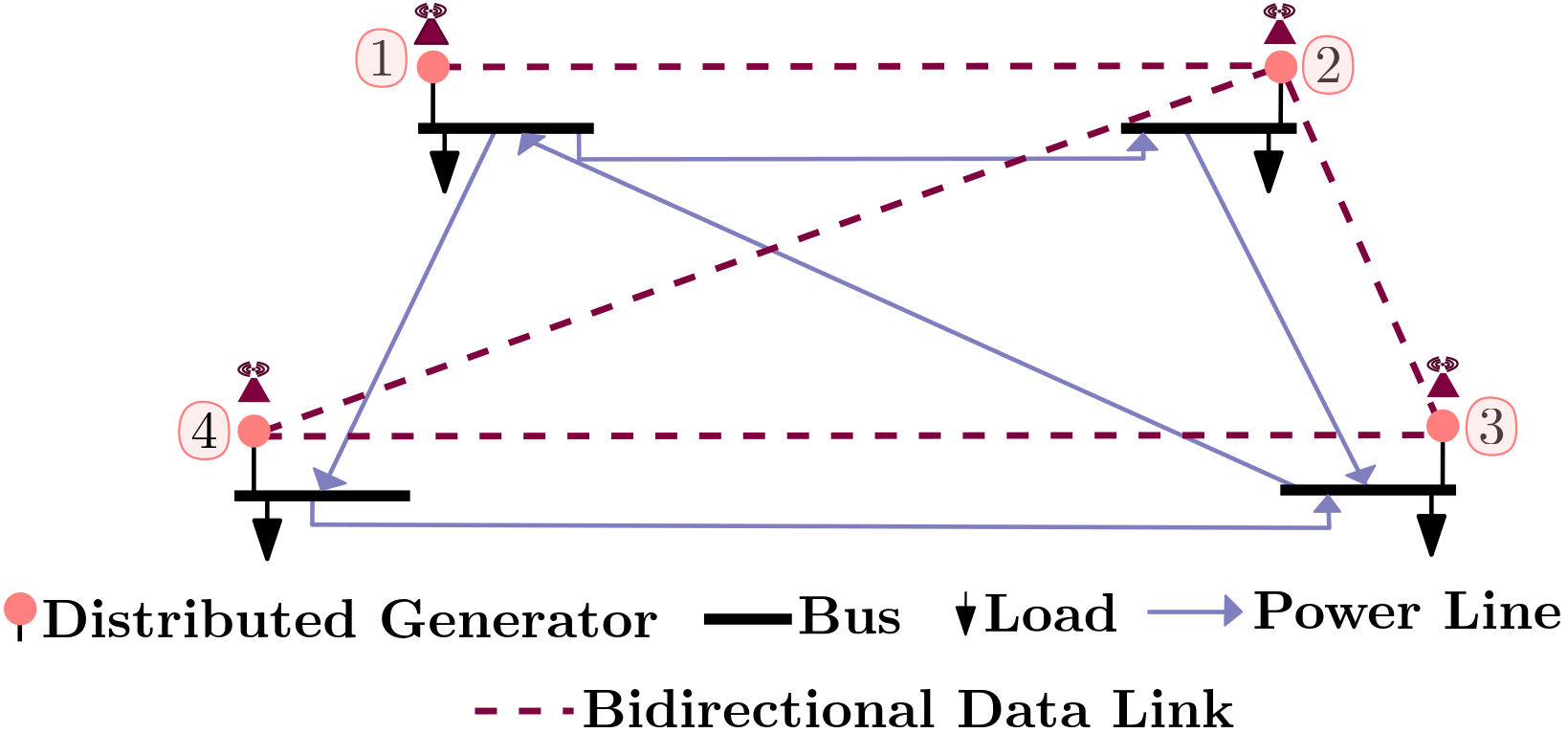}
    \captionsetup{font=small}
    \caption{Case-specific DC microgrid with four DGs.}
    \label{Image:Sim_Overview}
\end{figure}

The proposed control framework is tested by means of time-domain simulations in MATLAB/Simulink on a 48-volt DC network. The microgrid admits the dynamics in \eqref{eq:Physical-Layer}, powered by 4 DGs, interconnected electrically and through communication links according to the arbitrary interconnection patterns depicted in Fig. \ref{Image:Sim_Overview}. The specifications of the generators, loads, and power lines are given in Table \ref{Table_Spec}. For the initial system configurations, the selected control parameters—chosen to satisfy \emph{Assumption}~\ref{Asump_time_Const}—are listed below, with the time constants specified in seconds.  
\begin{align}\label{init_cont}
\begin{split}
    &\hspace{-0.32in}\tau = \mathrm{diag}(1e^{-3}),  \hspace{0.08in} \tau_p = \mathrm{diag}(0.01e^{-3}), \hspace{0.08in}  \tau_d = \mathrm{diag}(0.1e^{-3}), \\
    &\hspace{-0.32in}\mathcal{K}_v = \mathrm{diag}(V^*), \hspace{0.08in}\mathcal{B}_v= \mathrm{diag}(0), \hspace{0.08in} \alpha=V_\mathrm{max}, \hspace{0.08in} k = 10, \hspace{0.08in} b=5.
\end{split}
\end{align}
For the voltage containment objective, the maximum allowed voltage deviation from the nominal voltage, $V_n=48V$, are $5\%$; $V_{\mathrm{max}}=1.05$[p.u] and $V_{\mathrm{min}}=0.95$[p.u]. For the \emph{practical} proportional current sharing, a deviation of up to $\pm 5\%$ from the \emph{optimal} steady state operating point is permitted, at which \emph{ideal} proportional current sharing is achieved, characterized when the current integration error in \eqref{complete_sys_compact_v} converges to zero. Note that, although both $\mathcal{B}_v$ and $\mathcal{K}_v$ are diagonal matrices, their diagonal entries are selected uniformly across all DGs.
\subsection{Case Study 1: Control Performance under Stability Constraints}

\begin{figure*}
    \centering
    \includegraphics[width=\textwidth]{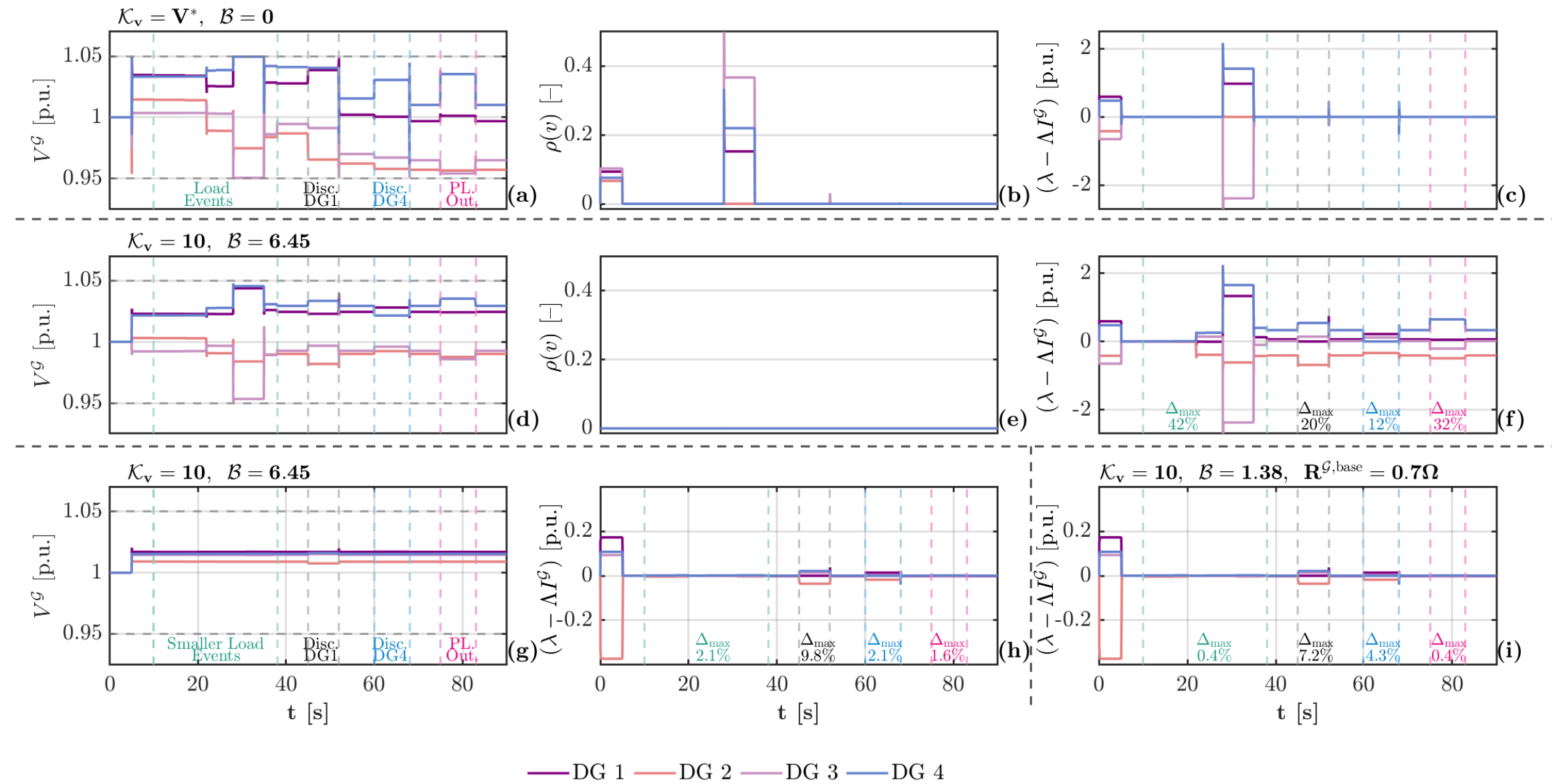}
    \captionsetup{font=small}
    \caption{Simulation results for Case Study 1.1 (a)-(c) (first row), Case Study 1.2 (d)-(f) (second row),  Case Study 1.3 (g)-(i) (third row); (a), (d), and (g) generators voltages; (b) and (e) \emph{non-permanent} leakage function; (c), (f), (h), and (i) integrations errors; Imposed load changes in Case Study 1 \& 2: t=5:  activation of distributed controller, $t=10: \mathrm{Inc.} G_1^\mathrm{cte} 45\%$, $t=17: \mathrm{Inc.} I_1^\mathrm{cte} 75\%$, $t=22: \mathrm{Inc.} G_2^\mathrm{cte} 500\%$ and $\mathrm{Inc.} I_1^\mathrm{cte} 800\%$, $t=24: \mathrm{Red.} I_4^\mathrm{cte} 35\%$, $t=28: \mathrm{Inc.} G_3^\mathrm{cte} 900\%$, $t=35: \mathrm{Red.} G_3^\mathrm{cte} 80\%$, $t=38: \mathrm{Red.} I_3^\mathrm{cte} 90\%$; Imposed load changes in Case Study 3: $t=5:$  activation of distributed controller, $t=10: \mathrm{Inc.} I_1^\mathrm{cte} 50\%$, $t=17: \mathrm{Red.} I_4^\mathrm{cte} 13\%$, $t=20: \mathrm{Red.} G_1^\mathrm{cte} 40\%$, $t=24: \mathrm{Inc.} G_2^\mathrm{cte} 5\%$, $t=31: \mathrm{Inc.} G_3^\mathrm{cte} 17\%$, $t=37: \mathrm{Red.} I_3^\mathrm{cte} 45\%$, $t=38: \mathrm{Red.} G_3^\mathrm{cte} 7\%$; Imposed disconnections in Case Study 1\&2\&3: $t\in[45,52]$ Disc. DG1, $t\in[60, 68]$ Disc. DG4, $t\in[75,83]$ Disc. PL between load 3 and 4.}
    \label{Fig:Simumations_X9}
\end{figure*}

In the following case studies, we aim to select a time-scale separation that \emph{practically} complies with the derived theoretical stability requirements via SPT. We then identify an acceptable controller tuning that guarantees system convergence to a \emph{near}-optimal steady state, ensuring both voltage containment and \emph{practical} proportional current sharing. Moreover, we impose various system events to evaluate the controller performance under both minor and more extreme load variations, as well as testing the robustness and stable plug-and-play capability under various network reconfiguration events--all imposed system events are defined in Fig.~\ref{Fig:Simumations_X9}. 

\textbf{Case Study 1.1: Initial System Configurations}\\
To evaluate whether the primary inner-loop controller consistently enforces strict voltage containment, the proposed control framework in \eqref{dynamics_inner_loop} and \eqref{dynamics_outer_loop} is tested on the case-specific system--parameters given in Table~\ref{Table_Spec} and \eqref{init_cont}--under extreme load variations, and without any stability guarantees. Thus, we select $\mathcal{B}_v=\mathrm{diag}(0)$. Moreover, to enforce voltage saturation scenarios, the load consumption of load 3 is increased accordingly; $G_3^\mathrm{cte}=G_3^{*\mathrm{cte}}/10$, and $I_3^\mathrm{cte}=10\times I_3^{*\mathrm{cte}}$, for $I^*$ and $G^*$ denoting the values listed in Table~\ref{Table_Spec}.
\begin{table}[!h]
    \centering 
    \captionsetup{font=small}
    \caption{Electrical Specifications for Case-Specific MG; resistance ($R$) and inductance ($L$) values are given in per unit (p.u) on a $0.15 \Omega,\; 300 \mu H$ base.}
    \label{Table_Spec}
    \begin{tabular} {|c|c|c|c|c|}
    \hline
    \multicolumn{5}{|c|}{Generator specifications;  $ i \in \mathcal{G}$}\\
    \Xhline{0.95pt}
    $I_i^\mathrm{rated}$ $[\mathrm{A}]$ & 12 & 4 & 8 & 8\\
    \hline
    $R_i^\mathcal{G} [\mathrm{p.u}]$,$L_i^\mathcal{G} [\mathrm{p.u}]$  & 0.5 & 0.4 & 0.55 & 0.6\\
    \Xhline{0.95pt}
    \multicolumn{5}{|c|}{Load specifications; $k \in \mathcal{N}$}\\
    \Xhline{0.95pt}
    $C_k^\mathcal{N} [\mathrm{F}]$  & \multicolumn{4}{c|}{$22 \times 10^{-3}$}\\
    \hline
    $1/G_k^\mathrm{cte} [\Omega]$ & 40 & 30 & 30 & 30 \\
    \hline
    $I_k^\mathrm{cte} [\mathrm{A}]$ & 1 & 1.2 & 0.8 & 1 \\
    \Xhline{0.95pt}
    \multicolumn{5}{|c|}{Power lines specifications; $j \in \mathcal{E}$}\\
    \Xhline{0.95pt}
    \multicolumn{5}{|c|}{
        \begin{tabular}{c|c|c|c|c|c}
        $R_j^\mathcal{E} [\mathrm{p.u}]$, $L_j^\mathcal{E} [\mathrm{p.u}]$ & 1 & 2 & 2 & 1 & 1 \\
        \end{tabular}} \\
    \hline
    \end{tabular}
\end{table}

The system response in Fig.~\ref{Fig:Simumations_X9}(a) shows that the DG voltages are always contained, and during the most significant load changes, some of the DG voltages reach their saturation limits, which activates their respective \emph{non-permanent} leakage $\rho(v)$, as depicted in Fig.~\ref{Fig:Simumations_X9}(b). Moreover, although the simulations in Fig.~\ref{Fig:Simumations_X9}(c) indicate \emph{near}-optimal steady state convergence, $\mathcal{B}_v$ remains inactive, and thus, the stability conditions in \emph{Theorem}~\ref{theorem_GES} are not satisfied. Hence, to guarantee stability under these system configurations, the elements \emph{permanent} leakage $\mathcal{B}_v$ must be chosen at least $349.5$ to satisfy \eqref{stab_condition}. However, this value violates \emph{Assumption}~\ref{Asump_time_Const}, as $(\tau_i/\mathcal{B}_i^v)=2.86\mu \ll \tau_i^p,\tau_i^d$ meaning that the inner-loop controller operates at a faster time-scale than the outer-loop controller.

\textbf{Case Study 1.2: Stable System Configurations}\\
Accordingly, to evaluate the DC MG under sufficient stability conditions, it is first necessary to carefully determine acceptable $\mathcal{B}_v$-values, following the guidelines in Section~\ref{sec:beta}. 

Initially, it may appear reasonable to minimize the elements of $\mathcal{B}_v$ by selecting the smallest feasible values of $\mathcal{K}_v$. However, Section~\ref{sec:beta} highlights that, while reducing the elements of $\mathcal{K}_v$ decreases the elements of $\mathcal{B}_v$, it simultaneously strengthens the effect of the leakage function in the steady state operation of the outer-loop controller, thereby diminishing \emph{ideal} proportional current-sharing performance. Consequently, a trade-off exists between selecting a sufficiently small $\mathcal{B}_v$ and maintaining a sufficiently large $\mathcal{K}_v$, such that the system converges to a steady state in which the integration error approaches zero or deviates by at most $\pm 5\%$. Accordingly, selecting $\mathcal{K}_v=\mathrm{diag}(10)$ results in $\mathcal{B}_v = \mathrm{diag}(6.45)$, which aligns with the required effective time constant of the integral controller according to \emph{Remark}~\ref{rem_tau_eff}.

\begin{Remark}\emph{(Effective Time Constant of the Integral Controller)}\label{rem_tau_eff}
Note that, given by Assumption~\ref{ass:tau}, the time constants are approximated by the ratio of inertia to damping, which arguably yields a conservative approximation of the true time constants. Moreover, since the diagonal entries of $\mathcal{B}_v$ are selected uniformly according to the maximum value of $K_4^\top {R}^{\mathcal{G}-1} K_4\,\omega'_{\max}$, the selected permanent leakage gain is governed by the worst-case DG. Consequently, this selection yields a more conservative time-scale separation for the remaining generators. Hence, the resulting rate of the integral controller is deemed sufficient to satisfy the stability condition, provided that the resulting time constant of each DG$_i$ satisfies $\tau_i/\mathcal{B}_i^v > \tau_i^d\gg \tau_i^p$. This approximation is further supported by Case Study~3, in which stability is guaranteed (from a small-signal perspective) under reduced time-scale separation requirements. 
\end{Remark}

The simulation results--depicted in Fig. \ref{Fig:Simumations_X9} (d)-(f)--then feature the system response of the same case-specific system, under similar loading conditions and system parameters as in the previous case study; however, under a controller tuning satisfying \emph{Theorem} \ref{theorem_GES} and \emph{Assumption} \ref{Asump_time_Const}. From a performance perspective, Fig. \ref{Fig:Simumations_X9}(d) illustrates that the voltages remain contained during all system events. However, compared to the previous case study--Fig. \ref{Fig:Simumations_X9}(a)-(b)--the voltages never reach their saturation limits due to the continuously active \emph{permanent} leakage, which significantly reduces the value of $v$, and thus the \emph{non-permanent} leakage remains deactivated throughout the simulations, as shown in Fig. \ref{Fig:Simumations_X9}(e). Although reducing the integrator gain $\mathcal{K}_v$ generally slows the integration process, and thus the rate of convergence, the increased value of $\mathcal{B}_v$ accelerates the convergence of $\dot v$, and due to the high communication frequencies, the system maintains its rapid convergence to steady state.
In regard to steady state accuracy, the active \emph{permanent} leakage--combined with extreme load variations--prevents convergence to a steady state for which \emph{ideal} proportional current sharing is satisfied. Specifically, the results in Fig. \ref{Fig:Simumations_X9}(f) indicate deviations of up to $42\%$, quantified by $\Delta_\mathrm{max}= \left[\frac{(\lambda_i-\Lambda_i I_i^\mathcal{G})}{\lambda_i}\right]_\mathrm{t}$, evaluated at each time instant $t$. Consequently, \emph{practical} proportional current sharing cannot be guaranteed under such extreme load variations. Despite this, the simulation results demonstrate stable operations and robust plug-and-play capability, maintaining system stability under significant load variations and topological changes in the electrical network.

\textbf{Case Study 1.3: Stable System Configurations and Practical Proportional Current Sharing}\\
Aiming to ensure \emph{practical} proportional current sharing, the system is simulated under more realistic load variations, while retaining the parameters specified in Table~\ref{Table_Spec} and \eqref{init_cont}, with  $\mathcal{K}_v=\mathrm{diag}(10)$ and $\mathcal{B}_v=\mathrm{diag}(6.45)$. Compared to the previous simulations--where load consumption was varied between a minimum of 35\% and a maximum of 900\%--the current study limits load changes to 50\%, reflecting more realistic conditions for a low-voltage MG. 

The voltages depicted in Fig. \ref{Fig:Simumations_X9}(g) are continuously contained, and thus, $\rho(v)$ remains inactive in a similar manner as in the previous case study. Fig.~\ref{Fig:Simumations_X9}(h) demonstrates that \emph{practical} proportional current sharing is achieved under the imposed load variations, power line disconnection, and disconnection of DG4. However, a deviation of $9.8\%$ is observed following the disconnection of DG1, which exceeds the selected threshold required to satisfy the \emph{practical} proportional current sharing objective. This behavior arises from the limited generation capacity of the considered low-voltage DC microgrid, as the disconnection of DG1 (i.e., the highest-rated generator) introduces a more substantial perturbation on the system operation, which has a more negative effect on the proportional current sharing objective.

Finally, following the supportive strategy illustrated in Fig.~\ref{Fig-TA}--or including a virtual impedance loop per \emph{Remark}~\ref{rem_Virtual_Impedance}-- increasing the generator's base resistance to $0.7\Omega$ reduces the values of $\mathcal{B}_v$ to $1.38$. However, as depicted in Fig.~\ref{Fig:Simumations_X9}(i), the combination of elevated generator resistances and a reduced $\mathcal{B}_v$ does not necessarily improve the controller's ability to guarantee \emph{ideal} proportional current sharing at the converged steady state. This is evident as the maximum deviation is reduced during the load events and the disconnection of the power line and DG1. Conversely, it increases following the disconnection of DG4. Accordingly, the final controller tuning and the resulting value of $\mathcal{B}_v$ are inherently system-dependent, and the corresponding performance is expected to vary with the system configurations and loading conditions. 

Nevertheless, the results in Fig.~\ref{Fig:Simumations_X9}(g)-(i) confirm that plug-and-play stability is preserved and the voltages are always contained within its pre-specified limits without experiencing overshoot. Furthermore, the results show that the \emph{practical} proportional current sharing is achieved for disconnection events involving lower-rated generators and power lines, whereas performance degradation becomes more significant when the unit supplying the largest share of power is removed.
\begin{figure*}[t!]
    \makebox[\textwidth][l]{ 
        \includegraphics[width=0.95\textwidth]{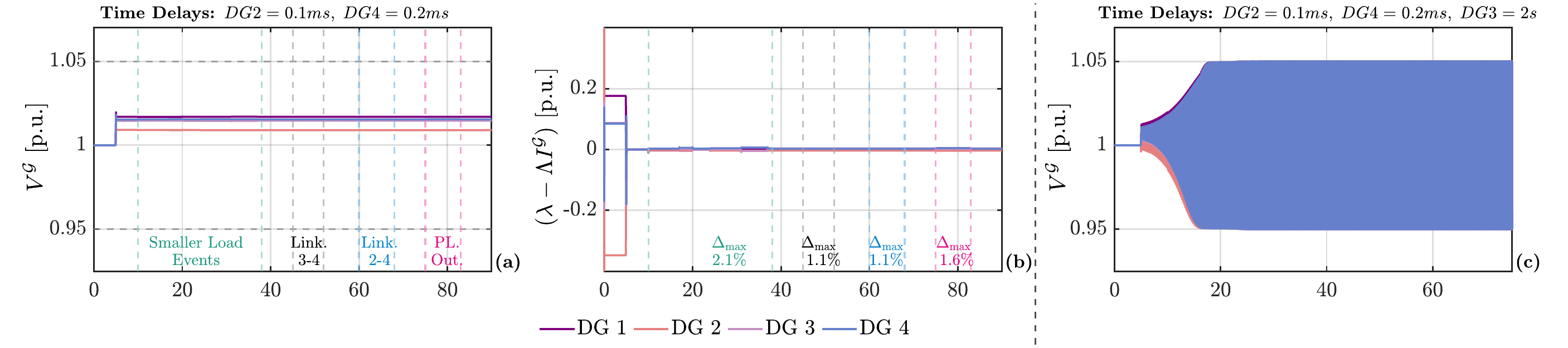}
    }
    \captionsetup{font=small}
    \caption{Case-specific microgrid under time delays; (a)(b) system response under time delays that respect the time-scale separation: (a) generator voltages, (b) integration error, (c) generator voltages under time delays that violate the time-scale separation.}
    \label{Fig_TD}
\end{figure*}
\begin{figure*}[!t]
    \centering
    \captionsetup{font=small}
    \includegraphics[width=0.95\textwidth]{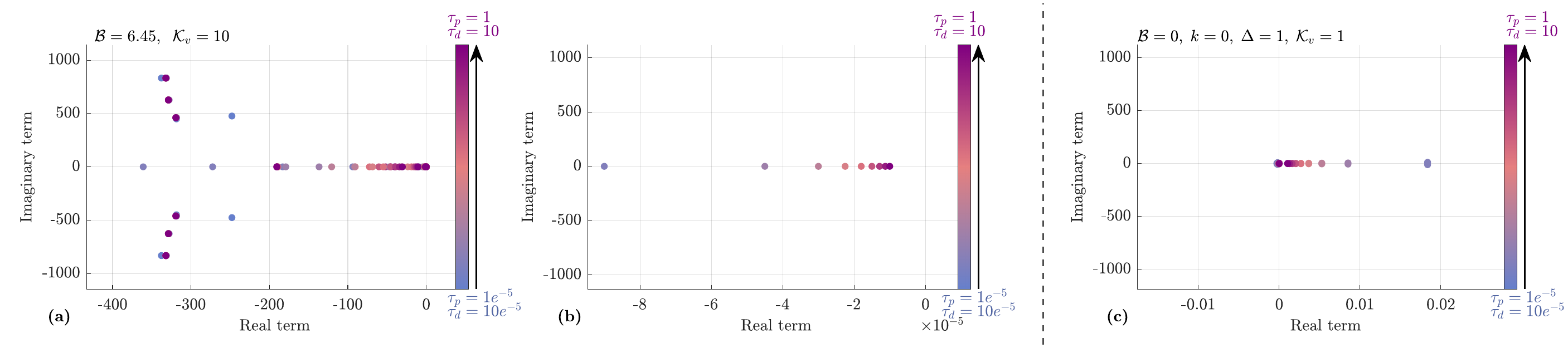}
    \caption{Eigenvalues of the case-specific linearized MG model for a parametric sweep of $\tau_p, \tau_d$; (a) and (b) eigenvalues under acceptable control parameters, (b) detailed plot around zero, (c) eigenvalues under unstable tuning conditions (detailed plot around zero).}
    \label{Fig:eig}
\end{figure*}
\subsection{Case Study 2: Control Performance under 
Communication Disruptions}
Given that the G.E.S. certificate relies on high communication rates, the resulting stability guarantees become sensitive to communication disruptions such as delays, packet losses, link failures, and improper tuning, which may violate the required time-scale separation. Accordingly, the objective of this case study is to assess how such communication disruptions may affect the underlying stability conditions and the controller's ability to consistently achieve consensus and \emph{practical} proportional current sharing. The same case-specific DC MG and loading conditions as in Case Study 1.3 are considered with $\mathcal{K}_v=\mathrm{diag}(6.45)$. As suggested in \cite{TD}, imposing constant time delays in the communication links can be interpreted as a proxy for communication packet losses. Hence, the outer-loop controller is subjected to imposed time delays, and instead of disconnecting DG1 and DG4 from the electrical network as in Case Studies 1, communication link failures are introduced on the DG2–DG3 and DG2–DG4 links over the respective time-intervals.
\begin{Remark}
    Note that, given by Assumption~\ref{ass_Laplacian}, the stability guarantees require a strongly connected communication graph, while overall consensus requires all active DGs to be virtually interconnected. Hence, under the topology in Fig. \ref{Image:Sim_Overview}, a failure in e.g. the DG1-DG2 communication link prevents DG1 from participating in the cooperative consensus formation process. However, a failure in the DG2–DG3 or DG2–DG4 communication links preserves the cooperative consensus formation among all active DGs while maintaining stability due to the remaining strongly connected graph.
\end{Remark} 
In the primary scenario--depicted in Fig. \ref{Fig_TD} (a), (b)--DG2 and DG4 respectively communicate with a time delay of $0.1$ms, and $0.2$ms--selected to respect the required communication rates. Hence, Fig.~\ref{Fig_TD}(a) and (b) illustrate that, despite the imposed time delays, the voltages remain bounded and the system converges to a steady state, with the maximum deviation from \emph{ideal} current sharing remaining consistent with the results reported in Fig.~\ref{Fig:Simumations_X9}(h) during the load events. In addition, the communication link failures induce negligible deviation from \emph{ideal} proportional current-sharing performance, with a maximum deviation of $1.1\%$ due to the remaining strongly connected communication network.

However, in Fig.~\ref{Fig_TD}(c), DG3 operates under a $2$ms time delay, leading to a violation of the required communication rate. The results indicate that this delay destabilizes the system, causing the generator voltages of all DGs to converge to a limit cycle due to their virtual inter-dependencies associated with the cooperative consensus formation. 

\subsection{Case Study 3: Small-Signal Stability Analysis}
Given that large-signal stability is preserved with substantial margins--characterized by the significant time-scale separation imposed in \emph{Assumption}~\ref{Asump_time_Const}--we perform a small-signal stability analysis to assess whether less conservative stability bounds exist. That is, we investigate
whether stable operation can be maintained under slower communication rates, thereby expanding the area of application of the results in this research. To this end, the closed-loop dynamics in \eqref{complete_sys_compact} are linearized around the equilibrium point of the considered case-specific DC microgrid.
We begin by noting that the outer-loop dynamics depend on the cyber-network topology defined by the Laplacian matrix, which inherently yields a zero eigenvalue due to its mathematical properties. Hence, to properly evaluate the closed-loop system's eigenvalues from a small-signal stability perspective, a small constant leakage, $\mathcal{B}_\zeta$, is introduced in the $\zeta$-coordinate in \eqref{complete_sys_compact_zeta} to add dissipation and shift the zero eigenvalue further into the left-half plane. That said, it is important to note that this leakage term affects the steady state operations of the system--particularly the controller's ability to determine a common consensus value of the dual communication state $\zeta$. As discussed in \emph{Remark}~\ref{lemma_Optimal_steady_state}, the \emph{optimal} equilibrium for \emph{ideal} proportional current sharing is defined when $\lambda_i = \lambda_j$ and $\zeta_i = \zeta_j$. 
Accordingly, we select $\mathcal{B}_\zeta=\mathrm{diag}(1e^{-4})$ as the maximum allowed value before hindering proportional current sharing. The resulting equilibrium values for $\lambda_i$ and $\zeta_i$ are detailed in Table~\ref{Table_equilibrium}, while retaining $\mathcal{B}_v=\mathrm{diag}(6.45)$ and $\mathcal{K}_v=\mathrm{diag}(10)$.  

\begin{table} 
    \centering 
    \captionsetup{font=small}
    \caption{Equilibrium of $\lambda$ and $\zeta$ with $\mathcal{B}_\zeta=\mathrm{diag}(1e^{-4})$} \label{Table_equilibrium}
    \resizebox{0.99\columnwidth}{!}{
    \begin{tabular} {|c|c|c|c|c|c|c|c|} 
    \hline
     $\lambda_1$ & $\lambda_2$ & $\lambda_3$ & $\lambda_4$ & $\zeta_1$ &$\zeta_2$ &$\zeta_3$ &$\zeta_4$\\ 
    \Xhline{1pt}
    0.3143 & 0.3143 & 0.3143 & 0.3143 & 0.0000 & 0.0000 & -0.0000 & -0.0000 \\
    \hline
    \end{tabular}
    }
\end{table}
To assess the impact of decreasing the communication rates, the eigenvalues of the case-specific DC MG are plotted for a parametric sweep of the communication time constants $\tau_p$ and $\tau_d$. The resulting eigenvalues are shown in Fig. \ref{Fig:eig} (a)-(b). From these figures, it is evident that reducing the communication rates brings the system closer to unstable operations as the eigenvalues tend towards positive real values. However, the eigenvalues never become positive, and the system is stable from a small-signal perspective. This implies that, from a small-signal perspective, the stability of the proposed closed-loop DC microgrid is not overly sensitive to the exact choice of time constants. In turn, this indicates that the SPT-based stability analysis yields a conservative stability certificate, with stability margins that extend beyond what is strictly necessary, which suggests that a less restrictive time-scale separation may be feasible in practice. However, under such relaxed time-scale separation requirements, neither \emph{practical scalability} nor large-signal stability can be guaranteed.

Furthermore, the small-signal stability assessment may be used to reinforce the findings of the large-signal stability analysis. Fig.~\ref{Fig:eig}(c) maps the eigenvalues of the closed-loop DC MG without careful tuning that guarantees overall stability; $\mathcal{B}_v=\mathrm{diag}(0)$, $k=0$, and decreasing the allowable voltage range ($\Delta=1$). The resulting plot depicts positive real eigenvalues, and the system is unstable. 
\begin{Remark}
    Note that in Fig.~\ref{Fig:eig}(c), eigenvalues shift toward negative real values when communication rates are significantly decreased, even with zero permanent leakage. This suggests a converse time-scale separation, where the outer-loop controller is slower than the rest of the system, removing the stability constraints on the permanent leakage-- which is addressed in \cite{Poppi_J2} and further explored in future work.
\end{Remark}
\section{Conclusion and Future Work}
\label{conc}
In conclusion, this paper derives a \emph{structurally scalable} global exponential stability (G.E.S.) certificate for a DC microgrid with nested nonlinear distributed control configurations. The closed-loop system is expressed as a singularly perturbed system, separating the outer-loop dynamics (secondary controller with inherent communication technologies), and the inner-loop dynamics (primary controller and electrical dynamics) by sufficient time-scales. 

Using Lyapunov stability theorems, this paper establishes G.E.S of the closed-loop DC microgrid, provided that the following two stability conditions are satisfied; (i) there \emph{exists} a sufficient time-scale separation between the nested control loops, and (ii) a \emph{permanent} leakage term of sufficiently large magnitude is implemented in the primary controller. Given that the required value of the \emph{permanent} leakage depends on both the controller tuning and some electrical specifications, practical guidelines are proposed to establish sufficient \emph{system configurations} that satisfies the derived G.E.S. condition for the worst-case scenario, under which the system is guaranteed to be \emph{practically scalable}, while aiming to enforce \emph{practical} proportional current sharing in steady state.

The derived stability conditions and the corresponding control performance are subsequently tested on a case-specific DC microgrid, included to characterize the required time-scale separation and determine controller tunings, and thus the acceptable and optimal \emph{permanent} leakage value. The sensitivity of the stability conditions {is also evaluated by introducing communication disruptions that may reduce or potentially violate the required time-scale separation. Additionally, a small-signal stability analysis is included to assess the feasibility of relaxing this time-scale separation, thereby enabling operation under less conservative conditions.

However, an important observation from the system simulations is that \emph{practical} proportional current sharing is achieved only under smaller (but arguably more realistic) load variations and for disconnection events involving DGs that do not represent the dominant generators in the considered low-voltage DC microgrid. Since the DC microgrid has limited generation capacity, the disconnection of the highest-rated DGs results in a more substantial perturbation of the system operating conditions, thereby introducing a more pronounced negative impact on the proportional current sharing objective. Nevertheless, the simulation results validate that the system rapidly converges to a stable steady state under all considered plug-and-play operations due to the fast communication rates, while the voltages remain contained within the predefined limits without experiencing overshoot.

Consequently, while the G.E.S. certificate is \emph{structurally scalable}, the resulting stability conditions depend on carefully selected \emph{system configurations}. Therefore, the practical scalability and general applicability of the results are primarily retained for expansions involving additional DGs (and loads) with similar or higher rated capacities than those considered in the case-specific DC microgrids. Nevertheless, future work should focus on generalizing some worst-case operating conditions that satisfy the stability conditions for expandable systems, considering a broader range of initial system configurations and DG connection/disconnection scenarios involving a wider range of power ratings, although this may adversely compromise proportional current sharing. Accordingly, the selection of the \emph{permanent} leakage should be guided by the intended engineering objective: either \emph{scalable} and stable operation across highly diverse networks, albeit with potentially reduced proportional power sharing performance, or stable and \emph{near-}optimal operation tailored to more case-specific systems. Moreover, it is important to note that, given that the theoretical large-signal stability guarantees rely on relatively fast communication rates, the current configuration is most suitable for smaller, localized systems in which such communication speeds can be practically achieved.

Finally, extending the stability analysis to account for communication imperfections, such as time-varying delays and packet drops, is left as a direction for future work.

\bibliographystyle{cas-model2-names}

\bibliography{refs}



\end{document}